\documentclass[%
 reprint,
 amsmath,amssymb,
 aps,
]{revtex4-2}

\usepackage{graphicx}% Include figure files
\usepackage{dcolumn}% Align table columns on decimal point
\usepackage{bm}% bold math
\begin{document}

\preprint{APS/123-QED}

\title{Discussing a transition from bounded to unbounded energy in a time-dependent billiard}

\author{$^1$Anne Kétri P. da Fonseca, $^1$Felipe Augusto O. Silveira, $^1$Célia M. Kuwana, $^2$Diego F. M. Oliveira, $^1$Edson D. Leonel}
\affiliation{{$^1$Departamento de Física, UNESP - Universidade Estadual Paulista, Avenida 24A, 1515, Bela Vista, Rio Claro, 13506-900, São Paulo, Brazil\\
$^2$School of Electrical Engineering and Computer Science - University of North Dakota, Grand Forks, Avenue Stop 8357, 58202, North Dakota, USA.}}

\date{\today}% It is always \today, today,
             %  but any date may be explicitly specified

\begin{abstract}
We revisit a time-dependent, oval-shaped billiard to investigate a phase transition from bounded to unbounded energy growth. In the static case, the phase space exhibits a mixed structure. The chaotic sea in the static scenario leads to average energy growth for a time-dependent boundary. However, inelastic collisions between the particle and the boundary limit this unbounded energy increase. This transition displays properties similar to continuous phase transitions in statistical mechanics, including scale invariance, interrelated critical exponents governed by scaling laws, and an order parameter/susceptibility approaching zero/infinity at the transition. Furthermore, the system exhibits an elementary excitation that promotes particle diffusion and lacks topological defects that provide modifications to the probability distribution function.
\end{abstract}

%\keywords{Suggested keywords}%Use showkeys class option if keyword
                              %display desired
\maketitle

%\tableofcontents

\section{Introduction}
\label{introduction}

Phase transitions have been observed in nature since the beginning of civilization and are studied across diverse fields and systems using a variety of theoretical and experimental approaches. For educational purposes, a commonly cited example in primary schools worldwide is the phase transition of water: when the temperature is lowered to $0~^oC$ at standard atmospheric pressure, water undergoes a phase transition from liquid to solid (ice). In the opposite direction, heating water to $100~^oC$ results in a phase transition from liquid to vapor. Both of these transitions are examples of first-order (discrete) phase transitions \cite{sethna2021statistical}.

Another example of a phase transition occurs in ferromagnetic materials such as iron \cite{khanna1991magnetic,grinstein1976ferromagnetic,gutzwiller1963effect}. In these materials, the ferromagnetic phase, characterized by ordered and aligned spins, experiences a change to a paramagnetic phase with predominantly disordered spins (non-magnetized) at a critical temperature $T_c$. This transition is classified as a second-order (continuous) phase transition, as indicated by a set of observables that characterize the behavior at the transition point. 

Numerous other phase transitions have been studied. For instance, certain materials undergo a transition to a superconducting state \cite{kiometzis1994critical,bianchi2002first,vojta2000quantum} at very low temperatures, exhibiting zero (or nearly zero) electrical resistance. This superconducting transition is also considered a second-order phase transition. At ultra-cold temperatures, atoms with specific magnetic properties can also condense into a single quantum state, forming a Bose-Einstein condensate \cite{wouters2010superfluidity,berman2008bose,chen2000unusual}. Furthermore, many materials experience structural phase transitions between solid phases. A notable example is the transformation of graphite into diamond under high pressure in carbon \cite{kvashnin2014phase,ma2012graphene,zaiser1997radiation}.

Other types of transitions are also observed in other scientific fields, such as nonlinear dynamics. For example, in the transition from integrability to non-integrability \cite{leonel2020characterization}, a system changes from regular behavior to mixed dynamics, where chaos emerges in phase space \cite{leonel2015dynamical}. In the integrable regime, the phase space is characterized exclusively by periodic or quasi-periodic motion, making the system's behavior predictable over time. Alternatively, in a mixed regime, phase space contains chaotic regions surrounded by periodic islands, and invariant tori may still exist. The chaotic dynamics in this context lead to diffusion, which becomes a crucial point of investigation. The chaotic diffusion exhibits scaling invariance \cite{leonel2005scaling}, described by a set of scaling hypotheses yielding a scaling law. This law provides an analytical relation between critical exponents near the phase transition \cite{pathria2011statistical}. Using scaling variables allows the diffusion curves for different control parameters to collapse onto a single and, hence, universal curve, confirming the presence of scaling invariance near the transition. Scaling invariance has also been observed in various systems, including the bouncing ball model \cite{oliveira2013some}, waveguides \cite{da2014escape}, social media networks \cite{oliveira2018scaling}, and billiard systems \cite{oliveira2012scaling, livorati2011family}.

A billiard is a dynamical system composed of a particle or a set of non-interacting particles undergoing specular collisions with a rigid boundary that confines them \cite{chernov2006chaotic}. Billiards are typically classified based on the shape of their boundary (e.g., circular \cite{berry1981regularity,bunimovich2005open}, elliptical \cite{koiller1996static,lenz2008tunable}, oval \cite{lopac2002chaotic,leonel2009fermi}, stadium \cite{tomsovic1993long,dettmann2009survival,lozej2018aspects}), which fundamentally influences the system's dynamics, resulting in fully integrable, chaotic, ergodic, or mixed behavior. These systems can be modeled by Hamiltonians of the form $H(x,p,t)=p^2/2m + V(x,t)$ where $V(x,t)=V_0(x)+V_1(x,t)$ and $V_0(x)$ is associated with time-dependent boundary variations, introduces non-integrability \cite{lichtenberg2013regular}. Notably, scaling behavior in these systems is frequently linked to diffusion processes, which serve as a valuable framework for understanding transitions and applications across various physical systems \cite{bohrer2020complex,xu2021global,lu2019modeling,gibert2019laplacian,chkhetiani2014detection}. The study of such diffusion mechanisms in billiards deepens our knowledge of such dynamics and provides an opportunity for potential applications in diverse fields.

This paper investigates a transition from limited to unlimited diffusion in a time-dependent oval billiard. The border is written in polar coordinates as $R(\theta,t)=1+\epsilon[1+\eta \cos(t)] \cos(p\theta)$, where $\epsilon$ controls the integrability of the system and $\eta$ defines the amplitude of the temporal dependence. Despite being an integrable system for $\epsilon=0$, a mixed phase space characterizes the dynamics for $\epsilon\ne 0$ \cite{chernov2006chaotic}. According to the Loskutov-Ryabov-Akinshin (LRA) conjecture \cite{loskutov2000properties}, the existence of chaos in the phase space is a sufficient condition for the occurrence of Fermi acceleration \cite{loskutov1999mechanism} (unlimited energy growth) when a time perturbation of the boundary is introduced. However, this phenomenon is not robust with the introduction of inelastic collisions with the boundary \cite{oliveira2010suppressing}.

We, therefore, revisit a time-dependent oval-like billiard to discuss a phase transition from bounded to unbounded diffusion. The system's dynamics is described by a discrete mapping $T(\theta_n, \alpha_n, V_n, t_n)$. Dissipation is introduced in the system via inelastic collision, allowing us to characterize the phase transition near the boundary between dissipative and conservative dynamics. In the dissipative case, the average velocity for an ensemble of particles approaches a stationary state, while the conservative dynamics lead to unbounded energy growth. We show that the behavior of the average velocity for an ensemble of particles can be described by the probability distribution function $P(V,n)$, which provides the probability of observing a given particle with a certain velocity at a specific time, thereby marking our original contribution to the problem. Furthermore, we aim to address four main questions: (i) what symmetry is broken at the transition? (ii) what is the order parameter, and how does its susceptibility behave at the transition? (iii) what are the topological defects? and (iv) how can we classify the elementary excitation that drives particle diffusion?

This paper is organized as follows: Section \ref{xsec2} describes the model for the time-dependent oval billiard and discusses its scaling properties. Sections \ref{xsec3} and \ref{xsec4} give the proposed probability distribution, such as the solution of the diffusion equation. The critical exponents are also presented. Section \ref{xsec5} presents the discussions and conclusions.

\section{The model and the mapping}
\label{xsec2}

\begin{figure}[t]
  \centering
    \centerline{(a)\includegraphics[width=1\linewidth]{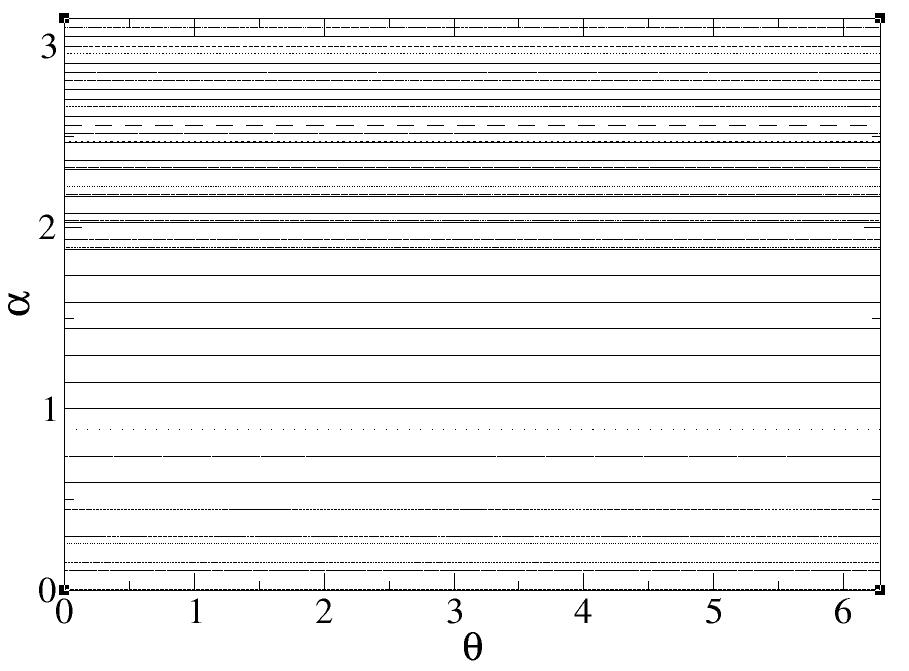}}
    \centerline{(b)\includegraphics[width=1\linewidth]{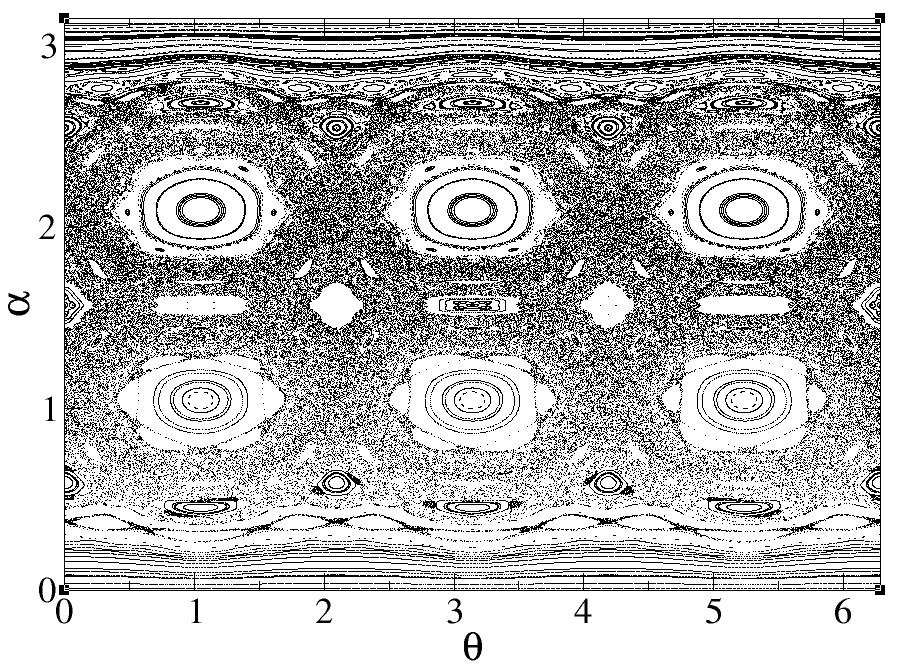}}    
  \caption{Plot of the phase space for: (a) $\epsilon=0$ and; (b) $\eta=0$ with $\epsilon=0.04$ and $p=3$.} \label{Fig1}
\end{figure}

The system consists of a particle or an ensemble of non-interacting particles confined to move within a boundary with radius written in polar coordinate as $R_b(\theta,t)=1+\epsilon[1+\eta \cos(t)] \cos(p\theta)$ where $\theta$ is the polar angle, $\epsilon$ is a parameter controlling the circle perturbation. For $\epsilon=0$, the system is integrable; the phase space is foliated, showing only periodic and quasi-periodic orbits. The parameter $\eta$ controls the amplitude of the time perturbation. For $\eta=0$ and $\epsilon\ne 0$, the phase space is mixed, containing both chaos, regular dynamics such as periodic islands and invariant spanning curves marking the existence of whispering gallery orbits for $\epsilon<\epsilon_c=\frac{1}{(1+p^2)}$ with $p$ denoting any integer number. Figure \ref{Fig1} shows a plot of the phase space for (a) $\epsilon=0$ and (b) $\eta=0$ with $\epsilon=0.04$ and $p=3$.

When $\eta\ne 0$, the mixed structure of the phase space, as shown in Fig. \ref{Fig1}(b), is destroyed. As claimed by the LRA conjecture \cite{loskutov2000properties}, the chaotic dynamics is a sufficient condition to produce unbounded diffusion for an ensemble of particles when a time perturbation to the boundary is introduced \cite{leonel2009fermi}. The dynamics of the system is described by a four-dimensional nonlinear mapping relating the dynamics of the impact $n$ with the $n+1$ via an operator $T$ for a set of dynamical variables $T(\theta_n, \alpha_n, V_n, t_n)=(\theta_{n+1}, \alpha_{n+1}, V_{n+1}, t_{n+1})$, where $\alpha$ gives the angle the trajectory of the particle does with a tangent line at the polar angle $\theta$. At the same time, $V$ denotes the particle's velocity, and $t$ is the time. To obtain the instant of the collision, we follow the trajectory of the particle for a time $t \ge t_n$ as
\begin{eqnarray}
X_p(t) &=& X(\theta_n ,t_n) + \vert \vec{V_n}\vert \cos(\alpha_n + \phi_n)(t-t_n), \\
Y_p(t) &
=& Y (\theta_n ,t_n) +  \vert \vec{V_n}\vert \sin(\alpha_n + \phi_n)(t-t_n).
\end{eqnarray}
The instant of the impact is obtained when $R_p(t) =\sqrt{X^2 (t) + Y^2 (t)}=R_b$. We use specular reflection law at a non-inertial referential frame of the boundary; hence, we have
\begin{eqnarray}
    \vec{V}_{n+1}\cdot \vec{T}_{n+1} =\vert \vec{V}_{n} \vert [\cos(\alpha_n+\phi_n)\cos(\phi_{n+1})] +\\
+\vert \vec{V}_{n} \vert [\sin(\alpha_n+\phi_n)\sin(\phi_{n+1})], \notag \\
    \vec{V}_{n+1}\cdot \vec{N}_{n+1}=-\vert \vec{V}_{n} \vert [-\cos(\alpha_n+\phi_n)\sin(\phi_{n+1})]- \notag  \\
    - \vert \vec{V}_{n} \vert [\sin(\alpha_n+\phi_n)\cos(\phi_{n+1})]  + 2\vec{V_b}(t_{n+1}) \cdot \vec{N}_{n+1}.
\end{eqnarray}
This leads the velocity of the particle in the $(n+1)^{th}$ collision to be written as 
\begin{equation}
\vert \vec{V}_{n+1}\vert = \sqrt{(\vec{V}_{n+1}\cdot \vec{T}_{n+1})^2 + (\vec{V}_{n+1}\cdot \vec{N}_{n+1})^2}.
\end{equation}
The new angle marking the trajectory of the particle is
\begin{equation}
\alpha_{n+1}=\arctan\left[\frac{\vec{V}_{n+1}\cdot \vec{N}_{n+1}}{ \vec{V}_{n+1}\cdot \vec{T}_{n+1}}\right].
\end{equation}

The phenomena we want to investigate are related to energy diffusion in time. The natural observable is then the square root of the averaged squared velocity, defined as
\begin{equation}
 \overline{V}=\sqrt{\frac{1}{M} \sum^M_{i=1} \frac{1}{n}  \sum^n_{j=1} V^2_{i,j}}.
\label{Eq_1}
\end{equation}
Two different averages were made in Eq. (\ref{Eq_1}): one over an ensemble of $M$ different initial conditions and the other over time, also called Birkov's average. 

For the conservative case, as discussed earlier \cite{lenz2008tunable,oliveira2011fermi,batistic2011fermi,oliveira2012scaling,batistic2014exponential}, the behavior of $V_{rms}$ can be summarized as: (i) for small initial velocities such as those with the order of the maximum velocity of the boundary, $V_{rms} \propto n^\beta$ with $\beta \approx 0.5$; (ii) for larger initial velocities a plateau is observed of constant $V_{rms}$ marked by $\overline{V}_{plat} \propto  V_0^\alpha$ is observed for $n \ll n_x$ and $\alpha \approx 1$; (iii) the changeover from the plateau to the growth regime is written as $n_x \propto V_0^z$, with $z\approx 2$. It is known this behavior can be described by a homogeneous and generalized function leading to a scaling law $z=\alpha/\beta$. With proper scaling variables, all $V_{rms}$ curves overlap onto a single and universal plot.

Our goal is to discuss the suppression of the unbounded growth, and we will do it by describing the probability density to observe a certain particle with a given velocity at an instant of time $n$. To do that, we consider the collisions of the particles with the boundary to be inelastic, introducing dissipation in the normal velocity component. Hence, the reflection law is given by
\begin{equation}
   \vec{V}_{n+1}\cdot \vec{N}_{n+1}=-\gamma \vec{V}_{n}  \cdot \vec{N}_{n+1} +  (1+\gamma)\vec{V_b}[t_{n+1} + Z(n)]  \cdot \vec{N}_{n+1}
\end{equation}
where $Z(n) \in [0,1]$ is a random number introduced to account for the stochasticity in the system, and $\gamma \in [0,1]$ is the restitution coefficient. For $\gamma=1$, the collisions are elastic, and all results for the non-dissipative case are obtained. 

\begin{figure}[t]
    \centerline{\includegraphics[width=1\linewidth]{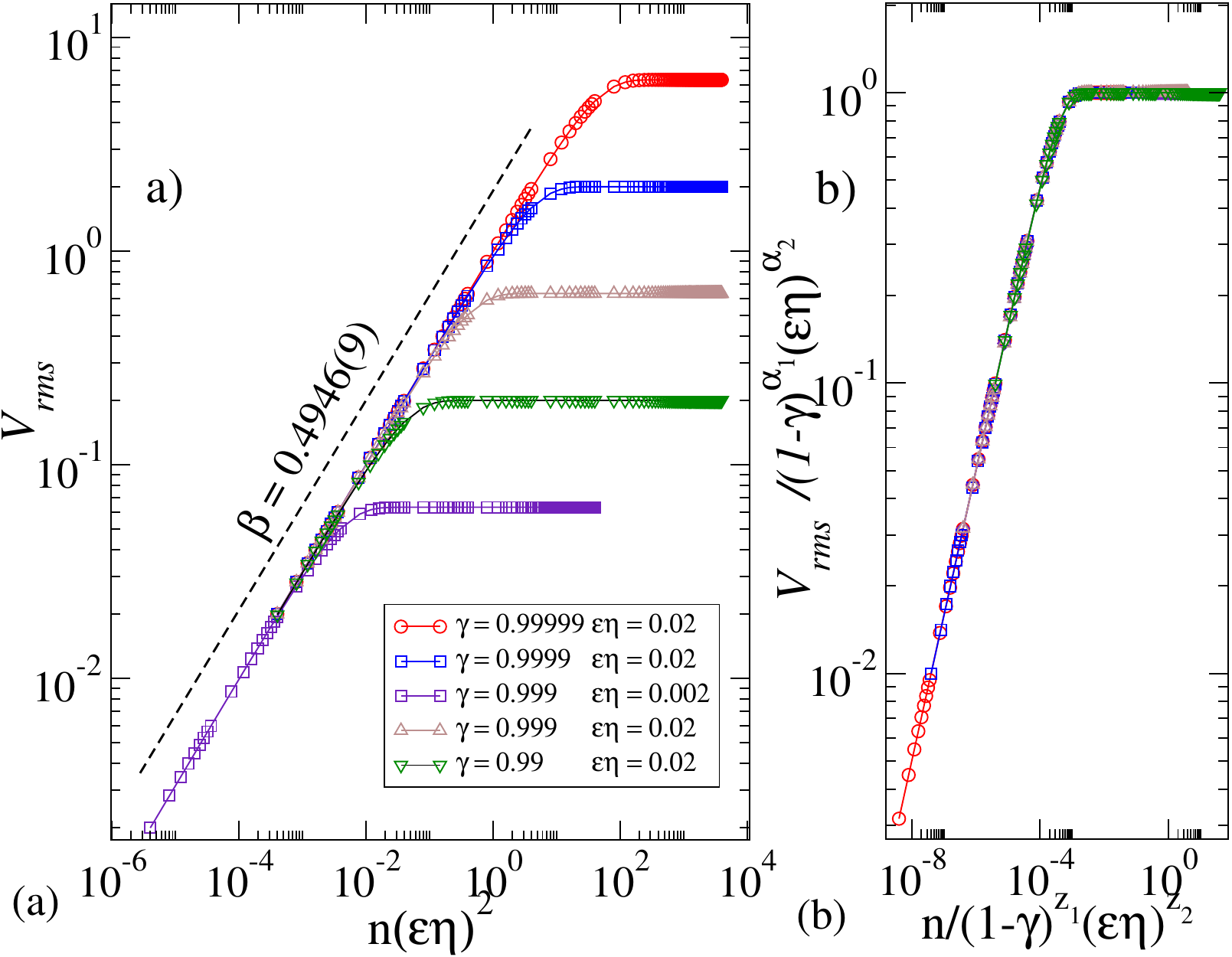}}
    % \centerline{\includegraphics[width=1\linewidth]{analliticavrmsBOM_b.eps}}    
\caption{(a) $V_{rms}$ $vs.$ $n(\epsilon \eta)^2$ constructed from the analytical expression for initial velocity $V_0 = 10^{-5}$ for different values of $\gamma$ and $\eta \epsilon$. (b) Overlap of the curves in \textit{a)} into a single universal curve using the the scaling transformations $V_{rms} \rightarrow V_{rms}/(1-\gamma)^{\alpha_1}(\epsilon\eta)^{\alpha2}$ and $n\rightarrow n/(1-\gamma)^{z_1}(\epsilon\eta)^{z_2}$.} 
\label{overlap}
\end{figure}

For $\gamma<1$, the fractional energy loss suppresses the FA. The system is described by a new set of scaling hypotheses and critical exponents: (i) for $n \ll n_x$ the growth regime is given by $V_{rms} \propto [(\eta\epsilon)^2n]^\beta$ with $\beta \approx 0.5$; (ii) for $n \gg n_x$ the saturation is described by $\overline{V}_{sat} \propto  (1-\gamma)^{\alpha_1}(\eta\epsilon)^{\alpha_2}$ where $\alpha_1 \approx -0.5$ and $\alpha_2 \approx 1$; (iii) the changeover from the growth regime to the plateau is written as $n_x \propto (1-\gamma)^{z_1}(\eta\epsilon)^{z_2}$, with $z_1 \approx -1$ and $z_2 \approx 0$. Figure \ref{overlap}(a) shows the behavior of the curves, as described above.

As in the conservative case, this behavior can be described by a homogeneous and generalized function leading to two scaling laws $z_1=\frac{\alpha_1}{\beta}$ and $z_2=\frac{\alpha_2}{\beta}-2$. 

The main point of the present paper is to describe the scaling result using the probability density function $P(V,n)$, which will be obtained by solving the diffusion equation, as discussed in the next section.

\section{Solution of the diffusion equation}
\label{xsec3}

Let us start the section by mentioning that the dynamics of the dissipative case lead to the suppression of the unbounded diffusion, hence imposing an upper limit for the velocity of the particles. We assume that the dynamics of the ensemble of particles behave like a particle moving randomly along the phase space, which turns out to be an important assumption for the application of the diffusion equation, which is written as
\begin{equation}
\frac{\partial P(V,n)}{\partial n} = D \frac{\partial^2P(V,n)}{\partial V^2}.
\end{equation}  
As mentioned, $P(V,n)$ gives the probability density to observe a given particle with a certain velocity $V$ at the instant $n$ while $D$ is the diffusion coefficient. It corresponds to how quickly particles can spread along the phase space over time. It quantifies the rate at which particles disperse through the phase space due to random motion (in our case, produced by equivalent chaotic dynamics). It then explains how fast the particles diffuse and furnishes an underlying mechanism producing the spread. Moreover, it reflects the tendency of particles to move from regions of higher concentration to areas of lower concentration. Therefore, a higher diffusion coefficient yields a faster spreading, while a lower one denotes a slower movement. As we will discuss shortly, the diffusion coefficient \( D \) is not constant in this investigation. It depends on the control parameters and time \( n \). However, its variation is sufficiently slow that it can be considered constant for small variations in \( n \). 

To solve the diffusion equation, we impose the following boundary conditions: (i) $P(V,n)|_{V\rightarrow0} =  P(V,n)|_{V\rightarrow\infty} = 0$ with a specific initial condition of the type; (ii) $P(V, 0) = \delta(V-V_0)$, ensuring that at $n=0$ all particles are starting from the same initial velocity distributed in $M$ different initial conditions uniformly distributed over $\alpha$, $\theta$ and $t$. Using the image method formalism combined with the distribution for a semi-infinite line \cite{balakrishnan2008elements}, we end up with the solution
\begin{equation}
    P(V,n)= \frac{\tau}{\sqrt{4\pi D n}} \left[ e^{ \left( \frac{-(V-V_0)^2}{4Dn} \right)} - e^{\left( \frac{-(V+V_0)^2}{4Dn} \right)} \right],
\end{equation}
with $\tau=erf\left(V_0/\sqrt{4Dn}\right)$ being a constant that emerges from the condition that $\int^\infty_{-\infty}P(V,n)dV=1$, \textit{i.e}, $P(V,n)$ must be normalized. 

\begin{figure}[t]
\centering
    \centerline{\includegraphics[width=0.91\linewidth]{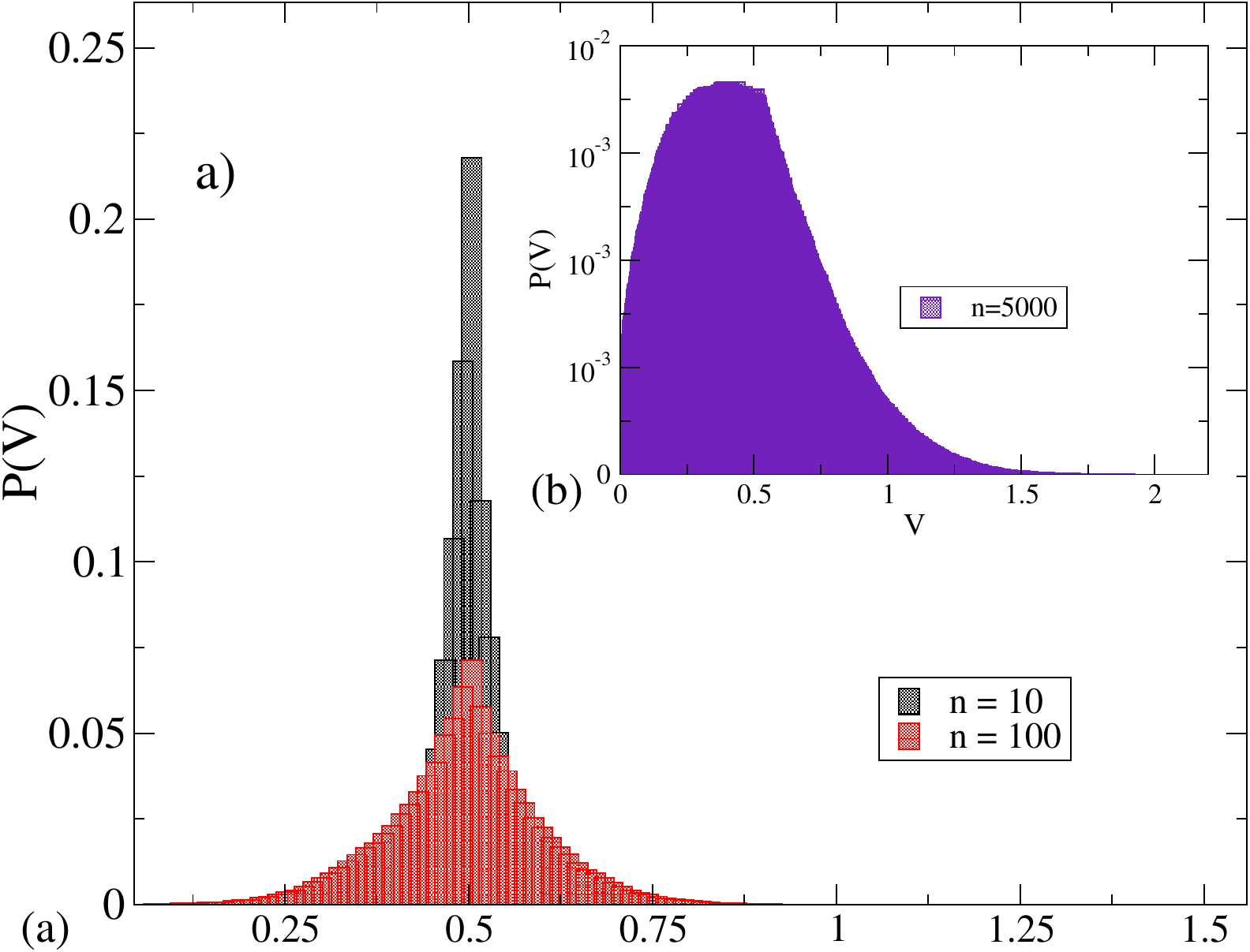}}
    \centerline{\includegraphics[width=0.9\linewidth]{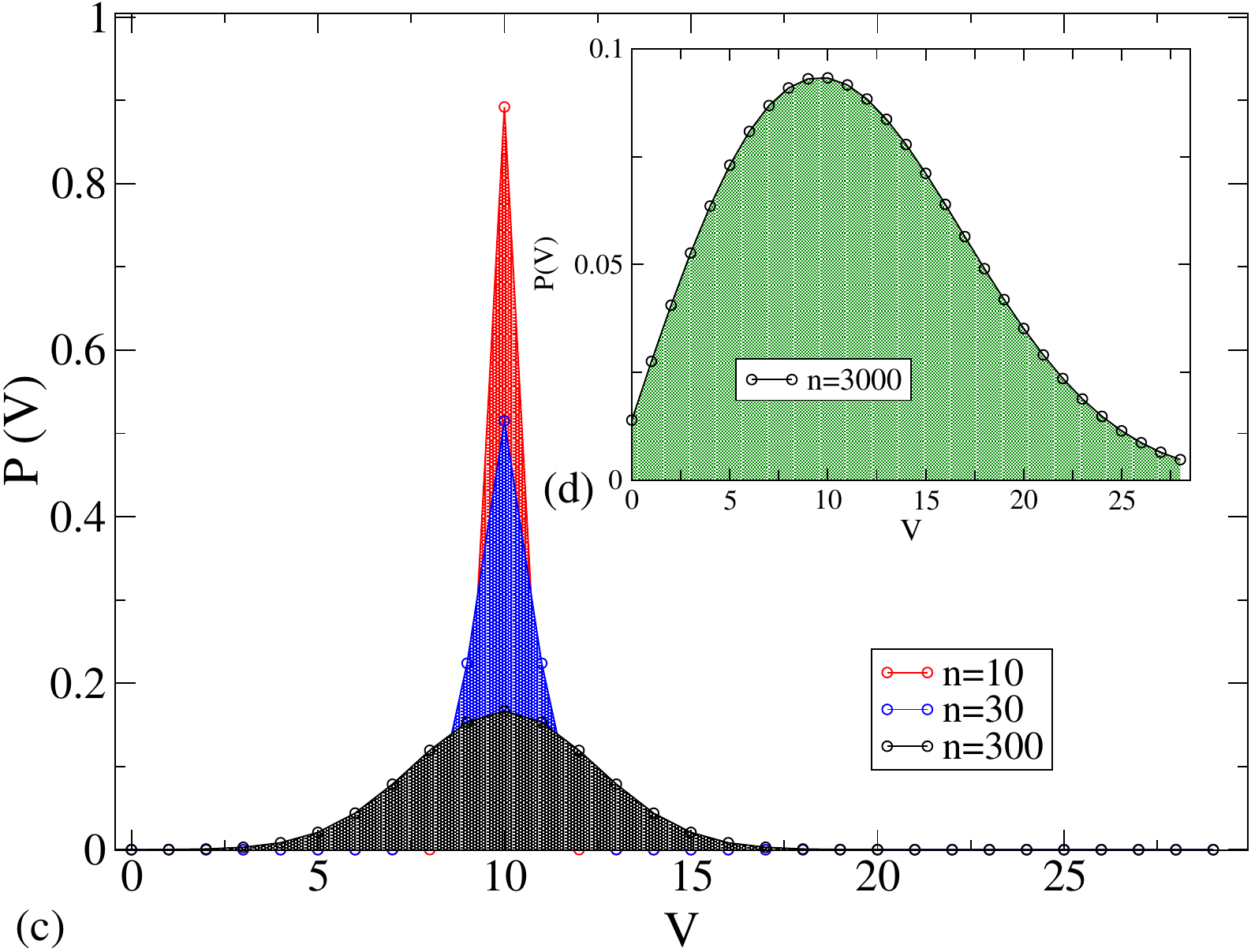}}    
\caption{(a) Plot of $P(V)$ for an ensemble of $M=3500$ particles in a dissipative time dependent oval billiard for the parameters $\eta=0.2$, $\epsilon=0.1$, $\gamma=0.999$, $p=3$ and $V_0=0.5$ after $n=10$, $100$ and (b) $5000$ collisions. (c) A plot of the analytical distribution was constructed using equation mapping for $D=0.01$,$V_0=10$, $n=10$, $30$, $300$, and (d) $3000$ collisions. The curves are rather similar to each other.} \label{fig:AB}
\end{figure}

A comparison between the phenomenological and the analytical distribution proposed can be seen in Fig. \ref{fig:AB} (a-d) for the initial velocity $V_0=0.5$ and parameters $\eta\epsilon=0.2$, $\gamma=0.999$ and $p=3$.

Obtaining the expression for $P(V,n)$ allows to write  $ \overline{V^2}(n)$ through the integration $ \overline{V^2}(n)=\int^{\infty}_{0}V^2 P(V,n)dV$, which gives
\begin{gather}
\overline{V^2}(n)= V_0^2 + 2Dn + V_0\tau\sqrt{\frac{4Dn}{\pi}}e^{-V_0^2/4Dn} ;
\end{gather}
and subsequently $V_{rms}(n)=\sqrt{\langle\overline{V}^2(n)\rangle}$, therefore
\begin{equation}
V_{rms}= \left( V_0^2 + 2Dn + V_0\tau\sqrt{\frac{4Dn}{\pi}} e^{-V_0^2/4Dn}\right)^{1/2}.
\end{equation}

Let us now discuss the diffusion coefficient $D$ specifically. It can be obtained from the mean squared displacement of particles over time
\begin{equation}
D=\frac{\overline{V^2_{n+1}}-\overline{V^2_{n}}}{2}=\frac{(\gamma^2-1)\overline{V^2_{n}}}{4} + \frac{(1+\gamma)^2\eta^2\epsilon^2}{16},
\end{equation}
where we considered and average over the following variables $t \in [0,2\pi]$, $\theta \in [0,2\pi]$ e $\alpha \in [0,\pi]$ obtained directly from the nonlinear mapping. The average velocity then reads as
\begin{equation}
 \overline{V^2_{n+1}} = \frac{(1+\gamma^2)\overline{V^2_{n}}}{2} + \frac{(1+\gamma)^2\eta^2\epsilon^2}{8}.
\end{equation}

\begin{figure}[t]
\centering
\centerline{\includegraphics[width=0.95\linewidth]{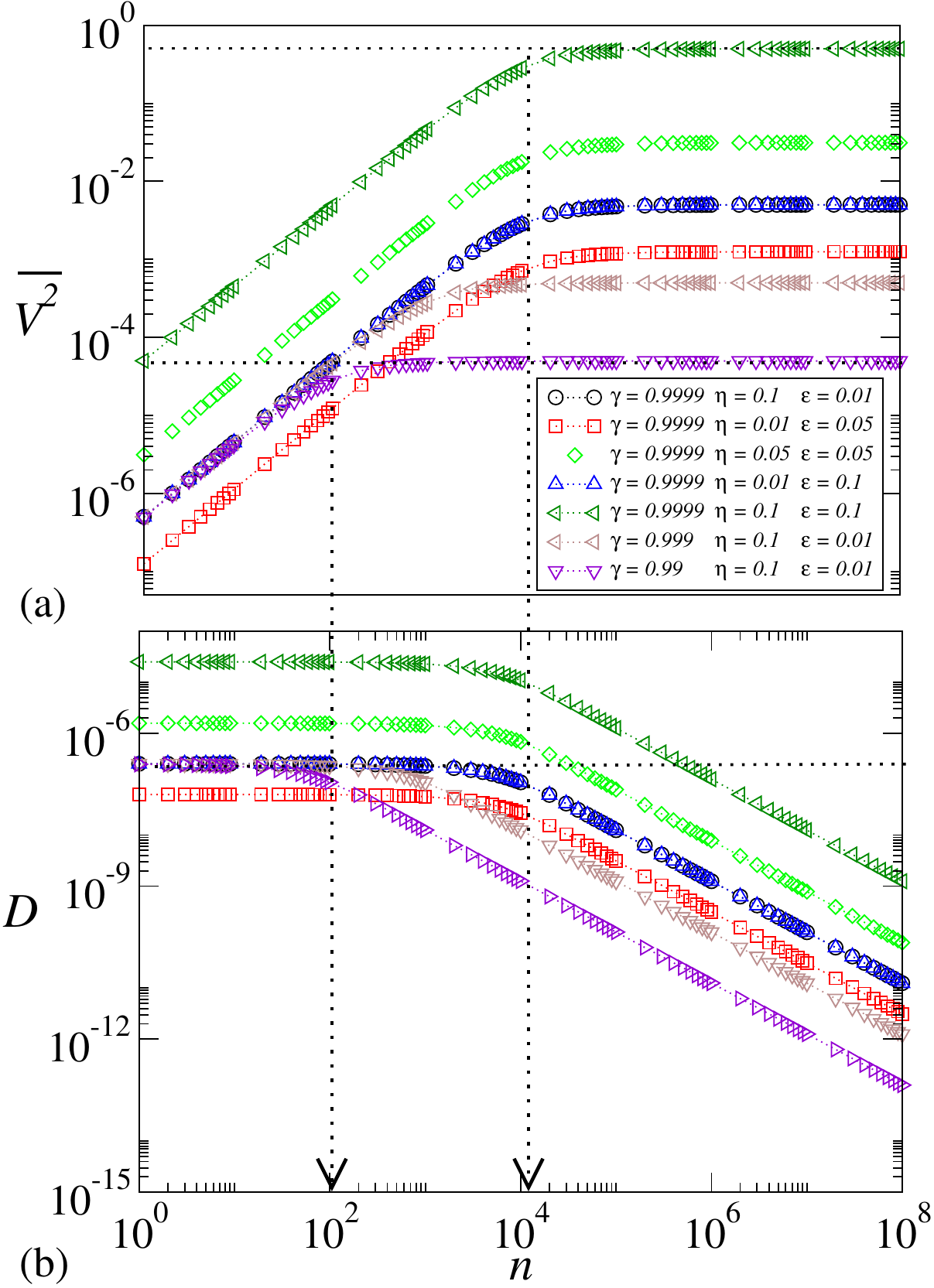}}
\caption{(a) Plot of $\overline{V^2}~vs.~n$ as given by Eq. (\ref{eq_v}) for different control, as shown in the figure. The behavior is remarkably similar to the previously presented in Fig. \ref{overlap}(a). (b) The plot of $D~vs.~n$ for the same control parameters of (a).}
\label{fig_new}
\end{figure}

Let us discuss the behavior of the average velocity. From the equations of the mapping and doing an average over an ensemble of initial conditions, we have that 
$\overline{V^2_{n+1}}-\overline{V^2_{n}}=\frac{\overline{V^2_{n+1}}-\overline{V^2_{n}}}{(n+1)-n} \simeq \frac{d\overline{V^2}}{dn}$. Integrating this expression in $n$, we end up with
\begin{eqnarray}
 \langle \overline{V^2}(n) \rangle &=& \frac{(1+\gamma)}{4(1-\gamma)}\eta^2\epsilon^2 + 
\notag \frac{1}{n+1} \left[\overline{V_0^2} - \frac{(1+\gamma)}{4(1-\gamma)}\eta^2\epsilon^2\right]  \nonumber \\
 &\times&\left[\frac{1-e^{\frac{(\gamma^2-1)(n+1)}{2}}}{1-e^{\frac{(\gamma^2-1)}{2}}}\right].
\label{eq_v}
\end{eqnarray}
Figure \ref{fig_new}(a) shows the behavior of $\overline{V^2}~vs.~n$ as given by Eq. (\ref{eq_v}) for different control parameters
while Fig. \ref{fig_new}(b) shows the behavior of $D~vs.~n$ for the same set of control parameters. 

It is easy to see from the figure that the curves $\overline{V^2}$ grow for a short time while the diffusion coefficient maintains a constant value, as previously assumed for the solution of the diffusion equation. It shows the particles are diffusing on the velocity axis as soon as the diffusion $D$ is constant. However, as the velocity increases, the dissipation suppresses the diffusion. As expected, the diffusion coefficient exhibits a changeover from the regime of continuous plateau to a regime of decay. The regime of growth of $\overline{V^2}$ is produced by the constant plateau of $D$ while the decrease of $D$ marks the saturation for $\overline{V^2}$, limiting the diffusion. The changeover from the regime of growth of $\overline{V^2}$ to the saturation coincides with the changeover from the domain of plateau for $D$ to its decay.

\begin{figure}[t]
\centering
\centerline{\includegraphics[width=0.95\linewidth]{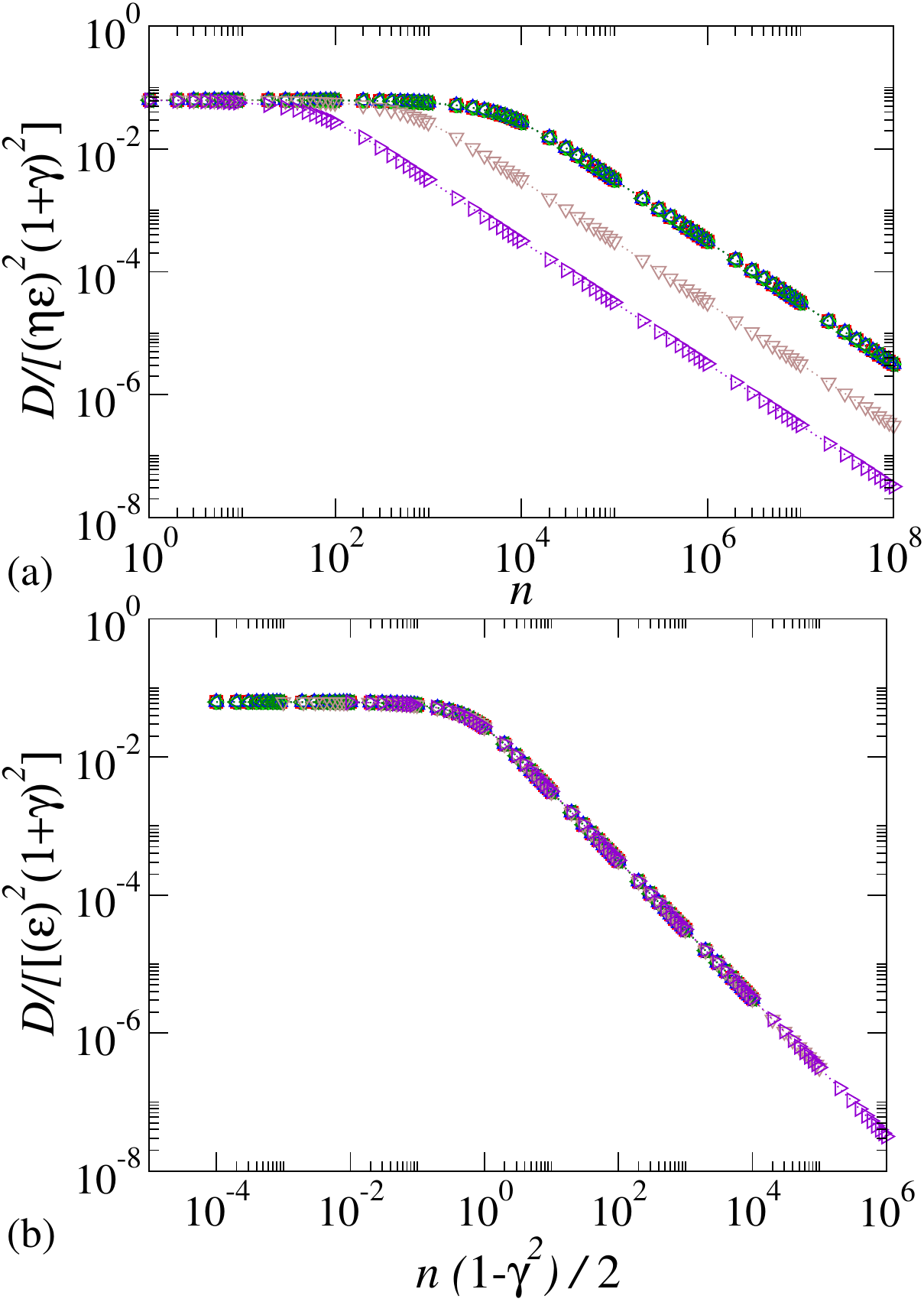}}
\caption{(a) Plot of $D~vs.~n$ for the transformation $D\rightarrow D/[(\eta\epsilon)^2(1+\gamma)^2]$. (b) shows a plot of $D~vs.~n$ for the transformations $D\rightarrow D/[(\eta\epsilon)^2(1+\gamma)^2]$ and $n\rightarrow n(1-\gamma^2)/2$.The control parameters are the same as those shown in Fig. \ref{fig_new}.}
\label{fig_new_overlap}
\end{figure}

As we notice, $\overline{V^2}$ behavior is scaling invariant. The diffusion coefficient $D$ is the same way. The diffusion coefficient behaves as a constant for a short time. After a crossover $n_x$, it bends towards a regime of decay. The scaling variables showing universality and hence scaling invariance for $D$ are: (i) $D\rightarrow D/[(\eta\epsilon)^2(1+\gamma)^2]$, as show an excellent overlap onto a single plot as seen in Fig. \ref{fig_new_overlap}(b). 
When a transformation $n\rightarrow n(1-\gamma^2)/2$ is applied to the horizontal axis, as shown in Fig. \ref{fig_new_overlap}(b), all curves of $D~vs.~n$ obtained for different control parameters overlap in a single universal plot, confirming the scaling invariance for the diffusion coefficient $D$.

Let us now discuss the implications of the results obtained for Eq. (\ref{eq_v}). Considering sufficiently long time, typically $\lim_{n\rightarrow\infty}$ we obtain that a saturation is reached for $\overline{V}_{sat} \propto  (1-\gamma)^{\alpha_1}(\eta\epsilon)^{\alpha_2}$. Comparing this result with the behavior of $V_{rms}$ as $n \rightarrow \infty$ given by 
\begin{equation}
V_{sat} \propto (1-\gamma)^{-1/2}\eta\epsilon,
\label{eqa}
\end{equation}
we found by an immediate analysis that $\alpha_1=-0.5$ and $\alpha_2=1.0$. Considering the limit of $n \ll n_x$ and assuming that $\gamma \rightarrow 1$, therefore close to the transition, we obtain
\begin{equation}
V_{rms}=\left[\frac{n(\gamma+1)^2}{8}\eta^2\epsilon^2\right]^{1/2}.
\label{eqb}
\end{equation}
As expected, a straightforward comparison with the scaling hypotheses $V_{rms} \propto n^{\beta}$ returns $\beta = 0.5$. Lastly, the behavior at the crossover can be written as equaling the Eqs. (\ref{eqa}) and (\ref{eqb}) leading to
\begin{equation}
n_x \propto \frac{1}{(1-\gamma)(1+\gamma)}.
\end{equation}
A comparison with the expression $n_x \propto (1-\gamma)^{z_1}(\eta\epsilon)^{z_2}$ returns the two last exponent: $z_1=-1$ and $z_2=0$. These results are also in agreement with the proposed scaling laws for the system, obtained through the homogeneous generalized function mentioned in Section 2;
\begin{gather}
z_1=\frac{\alpha_1}{\beta} = \frac{-0.5}{0.5} = -1, \\
z_2=\frac{\alpha_2}{\beta}- 2  = \frac{1}{0.5} - 2 = 0.
\end{gather}

\begin{figure}[t]
\centering
    \centerline{(a)\includegraphics[width=0.95\linewidth]{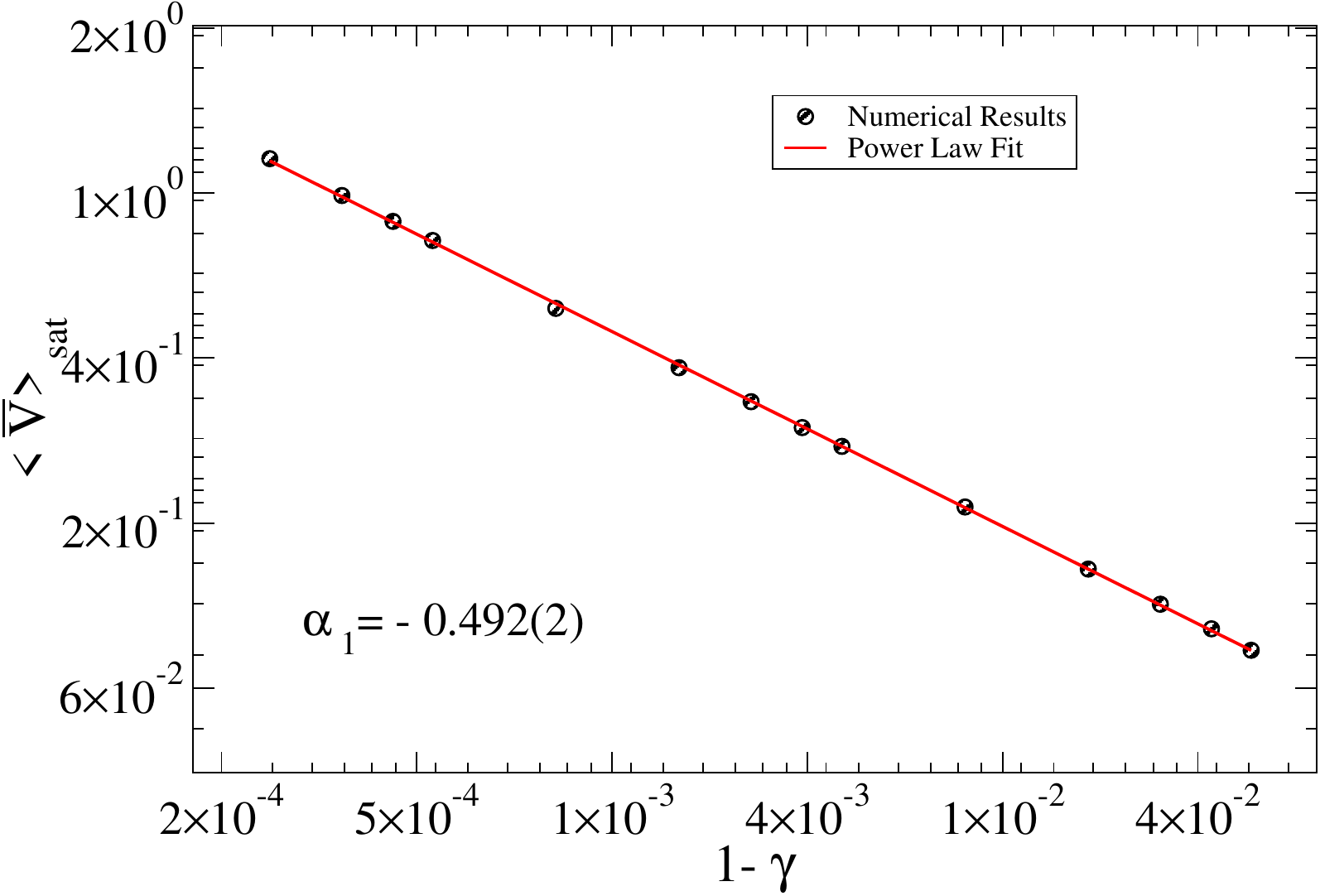}}
    \centerline{(b)\includegraphics[width=0.95\linewidth]{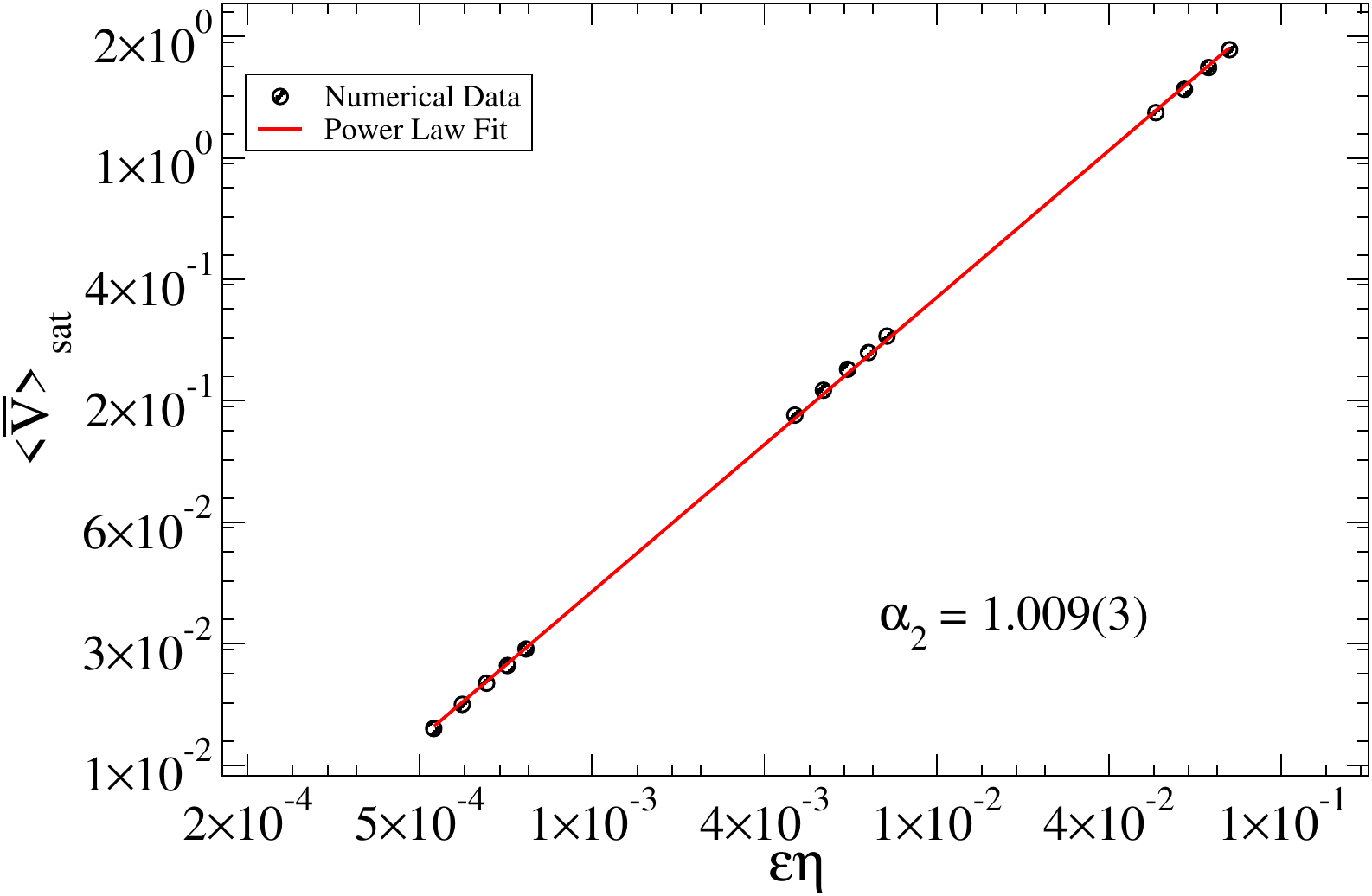}}    
\caption{(a) Plot of $V_{sat}$ $vs.$  $(1-\gamma)$. A power law fitting gives $\alpha_1=-0.492 (2)$. (b) Plot of $V_{sat}~vs.~\eta \epsilon$ with a power law fit giving $\alpha_2=1.009(3)$.}
\label{expoentes}
\end{figure}

These exponents can be obtained through the curves presented in Fig. \ref{overlap}(a). A power law fit from direct numerical simulations gives $\beta \approx 0.4946(9)$, while the values of $\alpha_1$ and $\alpha_2$ demand the construction of two new curves, as shown in Fig. \ref{expoentes}: (a) $V_{sat}$ $vs.$  $(1-\gamma)$ and; (b) $V_{sat}$ $vs.$ $ \eta \epsilon$ respectively, with the power law fit for those retrieving $\alpha_1=-0.492(2)$ and $\alpha_2=1.009(3)$, showing remarkable agreement with the established results for this system. 

Figure \ref{overlap}(b) presents the overlap of the curves shown in $(a)$ into a single and hence universal curve through the scaling transformations $V_{rms} \rightarrow V_{rms}/(1-\gamma)^{\alpha_1}(\epsilon\eta)^{\alpha2}$ and $n\rightarrow n/(1-\gamma)^{z_1}(\epsilon\eta)^{z_2}$.

\section{The phase transition}
\label{xsec4}

Let us now concentrate on discussing the elements characterizing the phase transition. A few essential characteristics are observed in a continuous phase transition \cite{sethna2021statistical} leading to a set of four main questions to be addressed: (i) Identify the symmetry breaking of the system at the transition; (ii) Define a suitable order parameter that approaches zero at the transition at the same time its susceptibility diverges in the same limit; (iii) Analyze the elementary excitations leading to the diffusion of particles and; (iv) Classify the topological defects of the system impacting on the transport of particles along of the phase space.

First, let us discuss the interpretation of an order parameter and how it affects the transition. Indeed, and as discussed in statistical mechanics, an order parameter is an observable that distinguishes between different phases of a system. It typically has a null value in a symmetric phase and a non-zero value in a non-symmetric phase. As we will see, for a second-order phase transition, the order parameter changes continuously from zero to a finite value as the system passes through the critical point (from conservative to dissipative). This behavior is confirmed by the scaling behavior observed for the saturation of the curves, shown for $V_{rms}$ and clearly described by the critical exponents. 

With this discussion in mind, we identify the quantity $\sigma = \frac{1}{V_{sat}}$ as a candidate to be an order parameter that attends the requirement of the classification of a second-order phase transition. Since $V_{sat}\propto (1-\gamma)^{-1/2}(\eta\epsilon)$ it is clear that $\sigma$ goes continuously to zero at the limit $\gamma\rightarrow 1$. Its susceptibility, marking the response of the order parameter to the variation of the dissipation parameter, accounts as
\begin{equation}
  \chi=  \frac{\partial}{\partial \gamma} \frac{1}{V_{sat}} =  \frac{\partial}{\partial \gamma} (1-\gamma)^{1/2}(\eta\epsilon)^{-1} = \frac{-(\eta\epsilon)^{-1}}{2(1-\gamma)^{1/2}} ,
  \end{equation}
diverges in the limit of vanishing dissipation, giving clear evidence of a second-order phase transition.

Discussing the order parameter allows us to identify the symmetry at the transition. For the conservative case where inelastic collisions are not present, and the unbounded growth of energy is observed, the average velocity has a behavior of the $V_{rms}\propto[(\eta\epsilon)^2n]^\beta$, with $\beta\approx 0.5$ leading the dynamics to show a symmetric phase for the regime of growth, i.e., a power law growth. In the limit of sufficiently large $n$, there is no bound for $V_{sat}$, making the order parameter converge continuously and smoothly to zero, marking a symmetric phase. At the same time, its susceptibility diverges. The scenario changes when dissipation is present. Instead of a regime of unbounded growth marked by a power law, the average velocity is described as a homogeneous and generalized function with a scaling law connecting critical exponents. For large enough $n$, a regime of saturation is observed, turning the order parameter to be a finite number so its susceptibility. There is a clear symmetry break of the behavior of $V_{rms}$ with an unbounded growth for the conservative case to the bounded growth for the dissipative case.

Elementary excitation is a term based on dynamics that are responsible for producing particle diffusion. For the case of $\epsilon=\eta=0$, the dynamics happen in a circular boundary shape. There is no boundary oscillation, and the dynamics is completely integrable. For $\epsilon\ne 0$ and $\eta=0$, the phase space shows a mixed structure with chaos circumventing periodic islands and invariant spanning curves. For the case of $\eta\ne 0$, the boundary is now time-dependent. The collisions of the particle with the border can now exchange energy and momentum, leading to diffusion in the velocity axis. The amount of velocity change upon collision is determined by the amplitude of the moving wall velocity, therefore a quantity that depends on the $V_b={\frac{dR_b(t)}{dt}}\propto\eta\epsilon$. We thus note the elementary excitation responsible for a velocity change and hence to the diffusion is the own amplitude of the moving wall velocity. Since it is the pre-factor of a periodic function cosine, its squared average is $V_a=\frac{\eta\epsilon}{\sqrt{2}}$. 

As a final point, the topological defects are structures in the phase space responsible for affecting particle transport. In a Hamiltonian case, the phase space has periodic islands that may lead to sticky dynamics near them. Given an orbit confined near a periodic island at a sticky trajectory may affect the probability density and hence the average properties of the dynamics, the periodic islands are assumed as topological defects \cite{leonel2020characterization} and influence the dynamics locally. In the dissipative case, periodic islands are absent due to the dissipation. In a generic case, the periodic islands centered by elliptical fixed points turn into sinks, hence attractors, that affect the dynamics of the particles when tiny dissipation is present. Sinks are not observed for the control parameters considered in the paper, leading us to assume that topological defects do not exist for the parameter ranges considered in the present investigation.

\section{Summary and conclusions}
\label{xsec5}

We characterized a transition from a dissipative to a non-dissipative dynamic for a time-dependent billiard. Our results conclude that the transition observed has properties that fit the transition at a continuous phase. The originality of the investigation lies in the analytical solution of the diffusion equation leading to the recovery of phenomenological results already known in the literature. The discussion allowed us to identify an order parameter approaching zero at the transition, marked by a symmetric phase confirmed by unbounded energy growth for a non-symmetric phase where the order parameter has a finite value. The elementary excitation of the system was identified as the amplitude of the moving wall velocity, hence imposing a maximum change of energy upon collision, yielding the particle to diffuse in the velocity axis. Finally, the topological defects related to the so-called sinks (attracting fixed points) are not observed in the phase space for the range of control parameters considered in the present paper. So far, the transition from bounded to unbounded energy growth in a time-dependent billiard has characteristics resembling a continuous phase transition where scaling invariance is present with scaling laws and critical exponents defining the criticality near the transition.

\section*{Acknowledgements}
A.K.P.F. acknowledges FAPESP $(2020/07219-1)$ for financial support. F.A.O.S. acknowledges CAPES (No. $88887.670331/2022-00)$ for financial support. CMK thanks to CAPES for support E.D.L. acknowledges support from Brazilian agencies CNPq (No. $301318/2019-0, 304398/2023-3)$ and FAPESP (No. $2019/14038-6$ and No. $2021/09519-5)$.

%\bibliography{references}% Produces the bibliography via BibTeX.

%apsrev4-2.bst 2019-01-14 (MD) hand-edited version of apsrev4-1.bst
%Control: key (0)
%Control: author (8) initials jnrlst
%Control: editor formatted (1) identically to author
%Control: production of article title (0) allowed
%Control: page (0) single
%Control: year (1) truncated
%Control: production of eprint (0) enabled
\begin{thebibliography}{45}%
\makeatletter
\providecommand \@ifxundefined [1]{%
 \@ifx{#1\undefined}
}%
\providecommand \@ifnum [1]{%
 \ifnum #1\expandafter \@firstoftwo
 \else \expandafter \@secondoftwo
 \fi
}%
\providecommand \@ifx [1]{%
 \ifx #1\expandafter \@firstoftwo
 \else \expandafter \@secondoftwo
 \fi
}%
\providecommand \natexlab [1]{#1}%
\providecommand \enquote  [1]{``#1''}%
\providecommand \bibnamefont  [1]{#1}%
\providecommand \bibfnamefont [1]{#1}%
\providecommand \citenamefont [1]{#1}%
\providecommand \href@noop [0]{\@secondoftwo}%
\providecommand \href [0]{\begingroup \@sanitize@url \@href}%
\providecommand \@href[1]{\@@startlink{#1}\@@href}%
\providecommand \@@href[1]{\endgroup#1\@@endlink}%
\providecommand \@sanitize@url [0]{\catcode `\\12\catcode `\$12\catcode `\&12\catcode `\#12\catcode `\^12\catcode `\_12\catcode `\%12\relax}%
\providecommand \@@startlink[1]{}%
\providecommand \@@endlink[0]{}%
\providecommand \url  [0]{\begingroup\@sanitize@url \@url }%
\providecommand \@url [1]{\endgroup\@href {#1}{\urlprefix }}%
\providecommand \urlprefix  [0]{URL }%
\providecommand \Eprint [0]{\href }%
\providecommand \doibase [0]{https://doi.org/}%
\providecommand \selectlanguage [0]{\@gobble}%
\providecommand \bibinfo  [0]{\@secondoftwo}%
\providecommand \bibfield  [0]{\@secondoftwo}%
\providecommand \translation [1]{[#1]}%
\providecommand \BibitemOpen [0]{}%
\providecommand \bibitemStop [0]{}%
\providecommand \bibitemNoStop [0]{.\EOS\space}%
\providecommand \EOS [0]{\spacefactor3000\relax}%
\providecommand \BibitemShut  [1]{\csname bibitem#1\endcsname}%
\let\auto@bib@innerbib\@empty
%</preamble>
\bibitem [{\citenamefont {Sethna}(2021)}]{sethna2021statistical}%
  \BibitemOpen
  \bibfield  {author} {\bibinfo {author} {\bibfnamefont {J.~P.}\ \bibnamefont {Sethna}},\ }\href@noop {} {\emph {\bibinfo {title} {Statistical mechanics: entropy, order parameters, and complexity}}},\ Vol.~\bibinfo {volume} {14}\ (\bibinfo  {publisher} {Oxford University Press, USA},\ \bibinfo {year} {2021})\BibitemShut {NoStop}%
\bibitem [{\citenamefont {Khanna}\ and\ \citenamefont {Linderoth}(1991)}]{khanna1991magnetic}%
  \BibitemOpen
  \bibfield  {author} {\bibinfo {author} {\bibfnamefont {S.}~\bibnamefont {Khanna}}\ and\ \bibinfo {author} {\bibfnamefont {S.}~\bibnamefont {Linderoth}},\ }\bibfield  {title} {\bibinfo {title} {Magnetic behavior of clusters of ferromagnetic transition metals},\ }\href@noop {} {\bibfield  {journal} {\bibinfo  {journal} {Physical review letters}\ }\textbf {\bibinfo {volume} {67}},\ \bibinfo {pages} {742} (\bibinfo {year} {1991})}\BibitemShut {NoStop}%
\bibitem [{\citenamefont {Grinstein}(1976)}]{grinstein1976ferromagnetic}%
  \BibitemOpen
  \bibfield  {author} {\bibinfo {author} {\bibfnamefont {G.}~\bibnamefont {Grinstein}},\ }\bibfield  {title} {\bibinfo {title} {Ferromagnetic phase transitions in random fields: the breakdown of scaling laws},\ }\href@noop {} {\bibfield  {journal} {\bibinfo  {journal} {Physical Review Letters}\ }\textbf {\bibinfo {volume} {37}},\ \bibinfo {pages} {944} (\bibinfo {year} {1976})}\BibitemShut {NoStop}%
\bibitem [{\citenamefont {Gutzwiller}(1963)}]{gutzwiller1963effect}%
  \BibitemOpen
  \bibfield  {author} {\bibinfo {author} {\bibfnamefont {M.~C.}\ \bibnamefont {Gutzwiller}},\ }\bibfield  {title} {\bibinfo {title} {Effect of correlation on the ferromagnetism of transition metals},\ }\href@noop {} {\bibfield  {journal} {\bibinfo  {journal} {Physical Review Letters}\ }\textbf {\bibinfo {volume} {10}},\ \bibinfo {pages} {159} (\bibinfo {year} {1963})}\BibitemShut {NoStop}%
\bibitem [{\citenamefont {Kiometzis}\ \emph {et~al.}(1994)\citenamefont {Kiometzis}, \citenamefont {Kleinert},\ and\ \citenamefont {Schakel}}]{kiometzis1994critical}%
  \BibitemOpen
  \bibfield  {author} {\bibinfo {author} {\bibfnamefont {M.}~\bibnamefont {Kiometzis}}, \bibinfo {author} {\bibfnamefont {H.}~\bibnamefont {Kleinert}},\ and\ \bibinfo {author} {\bibfnamefont {A.~M.}\ \bibnamefont {Schakel}},\ }\bibfield  {title} {\bibinfo {title} {Critical exponents of the superconducting phase transition},\ }\href@noop {} {\bibfield  {journal} {\bibinfo  {journal} {Physical review letters}\ }\textbf {\bibinfo {volume} {73}},\ \bibinfo {pages} {1975} (\bibinfo {year} {1994})}\BibitemShut {NoStop}%
\bibitem [{\citenamefont {Bianchi}\ \emph {et~al.}(2002)\citenamefont {Bianchi}, \citenamefont {Movshovich}, \citenamefont {Oeschler}, \citenamefont {Gegenwart}, \citenamefont {Steglich}, \citenamefont {Thompson}, \citenamefont {Pagliuso},\ and\ \citenamefont {Sarrao}}]{bianchi2002first}%
  \BibitemOpen
  \bibfield  {author} {\bibinfo {author} {\bibfnamefont {A.}~\bibnamefont {Bianchi}}, \bibinfo {author} {\bibfnamefont {R.}~\bibnamefont {Movshovich}}, \bibinfo {author} {\bibfnamefont {N.}~\bibnamefont {Oeschler}}, \bibinfo {author} {\bibfnamefont {P.}~\bibnamefont {Gegenwart}}, \bibinfo {author} {\bibfnamefont {F.}~\bibnamefont {Steglich}}, \bibinfo {author} {\bibfnamefont {J.~D.}\ \bibnamefont {Thompson}}, \bibinfo {author} {\bibfnamefont {.~f.~P.}\ \bibnamefont {Pagliuso}},\ and\ \bibinfo {author} {\bibfnamefont {J.~L.}\ \bibnamefont {Sarrao}},\ }\bibfield  {title} {\bibinfo {title} {First-order superconducting phase transition in c e c o i n 5},\ }\href@noop {} {\bibfield  {journal} {\bibinfo  {journal} {Physical review letters}\ }\textbf {\bibinfo {volume} {89}},\ \bibinfo {pages} {137002} (\bibinfo {year} {2002})}\BibitemShut {NoStop}%
\bibitem [{\citenamefont {Vojta}\ \emph {et~al.}(2000)\citenamefont {Vojta}, \citenamefont {Zhang},\ and\ \citenamefont {Sachdev}}]{vojta2000quantum}%
  \BibitemOpen
  \bibfield  {author} {\bibinfo {author} {\bibfnamefont {M.}~\bibnamefont {Vojta}}, \bibinfo {author} {\bibfnamefont {Y.}~\bibnamefont {Zhang}},\ and\ \bibinfo {author} {\bibfnamefont {S.}~\bibnamefont {Sachdev}},\ }\bibfield  {title} {\bibinfo {title} {Quantum phase transitions in d-wave superconductors},\ }\href@noop {} {\bibfield  {journal} {\bibinfo  {journal} {Physical review letters}\ }\textbf {\bibinfo {volume} {85}},\ \bibinfo {pages} {4940} (\bibinfo {year} {2000})}\BibitemShut {NoStop}%
\bibitem [{\citenamefont {Wouters}\ and\ \citenamefont {Carusotto}(2010)}]{wouters2010superfluidity}%
  \BibitemOpen
  \bibfield  {author} {\bibinfo {author} {\bibfnamefont {M.}~\bibnamefont {Wouters}}\ and\ \bibinfo {author} {\bibfnamefont {I.}~\bibnamefont {Carusotto}},\ }\bibfield  {title} {\bibinfo {title} {Superfluidity and critical velocities in nonequilibrium bose-einstein condensates},\ }\href@noop {} {\bibfield  {journal} {\bibinfo  {journal} {Physical review letters}\ }\textbf {\bibinfo {volume} {105}},\ \bibinfo {pages} {020602} (\bibinfo {year} {2010})}\BibitemShut {NoStop}%
\bibitem [{\citenamefont {Berman}\ \emph {et~al.}(2008)\citenamefont {Berman}, \citenamefont {Lozovik},\ and\ \citenamefont {Gumbs}}]{berman2008bose}%
  \BibitemOpen
  \bibfield  {author} {\bibinfo {author} {\bibfnamefont {O.~L.}\ \bibnamefont {Berman}}, \bibinfo {author} {\bibfnamefont {Y.~E.}\ \bibnamefont {Lozovik}},\ and\ \bibinfo {author} {\bibfnamefont {G.}~\bibnamefont {Gumbs}},\ }\bibfield  {title} {\bibinfo {title} {Bose-einstein condensation and superfluidity of magnetoexcitons in bilayer graphene},\ }\href@noop {} {\bibfield  {journal} {\bibinfo  {journal} {Physical Review B—Condensed Matter and Materials Physics}\ }\textbf {\bibinfo {volume} {77}},\ \bibinfo {pages} {155433} (\bibinfo {year} {2008})}\BibitemShut {NoStop}%
\bibitem [{\citenamefont {Chen}\ \emph {et~al.}(2000)\citenamefont {Chen}, \citenamefont {Kosztin},\ and\ \citenamefont {Levin}}]{chen2000unusual}%
  \BibitemOpen
  \bibfield  {author} {\bibinfo {author} {\bibfnamefont {Q.}~\bibnamefont {Chen}}, \bibinfo {author} {\bibfnamefont {I.}~\bibnamefont {Kosztin}},\ and\ \bibinfo {author} {\bibfnamefont {K.}~\bibnamefont {Levin}},\ }\bibfield  {title} {\bibinfo {title} {Unusual thermodynamical and transport signatures of the bcs to bose-einstein crossover scenario below t c},\ }\href@noop {} {\bibfield  {journal} {\bibinfo  {journal} {Physical Review Letters}\ }\textbf {\bibinfo {volume} {85}},\ \bibinfo {pages} {2801} (\bibinfo {year} {2000})}\BibitemShut {NoStop}%
\bibitem [{\citenamefont {Kvashnin}\ \emph {et~al.}(2014)\citenamefont {Kvashnin}, \citenamefont {Chernozatonskii}, \citenamefont {Yakobson},\ and\ \citenamefont {Sorokin}}]{kvashnin2014phase}%
  \BibitemOpen
  \bibfield  {author} {\bibinfo {author} {\bibfnamefont {A.~G.}\ \bibnamefont {Kvashnin}}, \bibinfo {author} {\bibfnamefont {L.~A.}\ \bibnamefont {Chernozatonskii}}, \bibinfo {author} {\bibfnamefont {B.~I.}\ \bibnamefont {Yakobson}},\ and\ \bibinfo {author} {\bibfnamefont {P.~B.}\ \bibnamefont {Sorokin}},\ }\bibfield  {title} {\bibinfo {title} {Phase diagram of quasi-two-dimensional carbon, from graphene to diamond},\ }\href@noop {} {\bibfield  {journal} {\bibinfo  {journal} {Nano Letters}\ }\textbf {\bibinfo {volume} {14}},\ \bibinfo {pages} {676} (\bibinfo {year} {2014})}\BibitemShut {NoStop}%
\bibitem [{\citenamefont {Ma}\ \emph {et~al.}(2012)\citenamefont {Ma}, \citenamefont {Dai}, \citenamefont {Guo},\ and\ \citenamefont {Huang}}]{ma2012graphene}%
  \BibitemOpen
  \bibfield  {author} {\bibinfo {author} {\bibfnamefont {Y.}~\bibnamefont {Ma}}, \bibinfo {author} {\bibfnamefont {Y.}~\bibnamefont {Dai}}, \bibinfo {author} {\bibfnamefont {M.}~\bibnamefont {Guo}},\ and\ \bibinfo {author} {\bibfnamefont {B.}~\bibnamefont {Huang}},\ }\bibfield  {title} {\bibinfo {title} {Graphene-diamond interface: Gap opening and electronic spin injection},\ }\href@noop {} {\bibfield  {journal} {\bibinfo  {journal} {Physical Review B—Condensed Matter and Materials Physics}\ }\textbf {\bibinfo {volume} {85}},\ \bibinfo {pages} {235448} (\bibinfo {year} {2012})}\BibitemShut {NoStop}%
\bibitem [{\citenamefont {Zaiser}\ and\ \citenamefont {Banhart}(1997)}]{zaiser1997radiation}%
  \BibitemOpen
  \bibfield  {author} {\bibinfo {author} {\bibfnamefont {M.}~\bibnamefont {Zaiser}}\ and\ \bibinfo {author} {\bibfnamefont {F.}~\bibnamefont {Banhart}},\ }\bibfield  {title} {\bibinfo {title} {Radiation-induced transformation of graphite to diamond},\ }\href@noop {} {\bibfield  {journal} {\bibinfo  {journal} {Physical review letters}\ }\textbf {\bibinfo {volume} {79}},\ \bibinfo {pages} {3680} (\bibinfo {year} {1997})}\BibitemShut {NoStop}%
\bibitem [{\citenamefont {Leonel}\ \emph {et~al.}(2020)\citenamefont {Leonel}, \citenamefont {Yoshida},\ and\ \citenamefont {de~Oliveira}}]{leonel2020characterization}%
  \BibitemOpen
  \bibfield  {author} {\bibinfo {author} {\bibfnamefont {E.~D.}\ \bibnamefont {Leonel}}, \bibinfo {author} {\bibfnamefont {M.}~\bibnamefont {Yoshida}},\ and\ \bibinfo {author} {\bibfnamefont {J.~A.}\ \bibnamefont {de~Oliveira}},\ }\bibfield  {title} {\bibinfo {title} {Characterization of a continuous phase transition in a chaotic system},\ }\href@noop {} {\bibfield  {journal} {\bibinfo  {journal} {Europhysics Letters}\ }\textbf {\bibinfo {volume} {131}},\ \bibinfo {pages} {20002} (\bibinfo {year} {2020})}\BibitemShut {NoStop}%
\bibitem [{\citenamefont {Leonel}\ \emph {et~al.}(2015)\citenamefont {Leonel}, \citenamefont {Penalva}, \citenamefont {Teixeira}, \citenamefont {Costa~Filho}, \citenamefont {Silva},\ and\ \citenamefont {De~Oliveira}}]{leonel2015dynamical}%
  \BibitemOpen
  \bibfield  {author} {\bibinfo {author} {\bibfnamefont {E.~D.}\ \bibnamefont {Leonel}}, \bibinfo {author} {\bibfnamefont {J.}~\bibnamefont {Penalva}}, \bibinfo {author} {\bibfnamefont {R.~M.}\ \bibnamefont {Teixeira}}, \bibinfo {author} {\bibfnamefont {R.~N.}\ \bibnamefont {Costa~Filho}}, \bibinfo {author} {\bibfnamefont {M.~R.}\ \bibnamefont {Silva}},\ and\ \bibinfo {author} {\bibfnamefont {J.~A.}\ \bibnamefont {De~Oliveira}},\ }\bibfield  {title} {\bibinfo {title} {A dynamical phase transition for a family of hamiltonian mappings: A phenomenological investigation to obtain the critical exponents},\ }\href@noop {} {\bibfield  {journal} {\bibinfo  {journal} {Physics Letters A}\ }\textbf {\bibinfo {volume} {379}},\ \bibinfo {pages} {1808} (\bibinfo {year} {2015})}\BibitemShut {NoStop}%
\bibitem [{\citenamefont {Leonel}\ and\ \citenamefont {McClintock}(2005)}]{leonel2005scaling}%
  \BibitemOpen
  \bibfield  {author} {\bibinfo {author} {\bibfnamefont {E.~D.}\ \bibnamefont {Leonel}}\ and\ \bibinfo {author} {\bibfnamefont {P.}~\bibnamefont {McClintock}},\ }\bibfield  {title} {\bibinfo {title} {Scaling properties for a classical particle in a time-dependent potential well},\ }\href@noop {} {\bibfield  {journal} {\bibinfo  {journal} {Chaos: An Interdisciplinary Journal of Nonlinear Science}\ }\textbf {\bibinfo {volume} {15}} (\bibinfo {year} {2005})}\BibitemShut {NoStop}%
\bibitem [{\citenamefont {Pathria}\ and\ \citenamefont {Beale}(2011)}]{pathria2011statistical}%
  \BibitemOpen
  \bibfield  {author} {\bibinfo {author} {\bibfnamefont {R.}~\bibnamefont {Pathria}}\ and\ \bibinfo {author} {\bibfnamefont {P.~D.}\ \bibnamefont {Beale}},\ }\href@noop {} {\bibinfo {title} {Statistical mechanics—third edition}} (\bibinfo {year} {2011})\BibitemShut {NoStop}%
\bibitem [{\citenamefont {Oliveira}\ and\ \citenamefont {Leonel}(2013)}]{oliveira2013some}%
  \BibitemOpen
  \bibfield  {author} {\bibinfo {author} {\bibfnamefont {D.~F.}\ \bibnamefont {Oliveira}}\ and\ \bibinfo {author} {\bibfnamefont {E.~D.}\ \bibnamefont {Leonel}},\ }\bibfield  {title} {\bibinfo {title} {Some dynamical properties of a classical dissipative bouncing ball model with two nonlinearities},\ }\href@noop {} {\bibfield  {journal} {\bibinfo  {journal} {Physica A: Statistical Mechanics and its Applications}\ }\textbf {\bibinfo {volume} {392}},\ \bibinfo {pages} {1762} (\bibinfo {year} {2013})}\BibitemShut {NoStop}%
\bibitem [{\citenamefont {da~Costa}\ \emph {et~al.}(2014)\citenamefont {da~Costa}, \citenamefont {Silva},\ and\ \citenamefont {Leonel}}]{da2014escape}%
  \BibitemOpen
  \bibfield  {author} {\bibinfo {author} {\bibfnamefont {D.~R.}\ \bibnamefont {da~Costa}}, \bibinfo {author} {\bibfnamefont {M.~R.}\ \bibnamefont {Silva}},\ and\ \bibinfo {author} {\bibfnamefont {E.~D.}\ \bibnamefont {Leonel}},\ }\bibfield  {title} {\bibinfo {title} {Escape beam statistics and dynamical properties for a periodically corrugated waveguide},\ }\href@noop {} {\bibfield  {journal} {\bibinfo  {journal} {Communications in Nonlinear Science and Numerical Simulation}\ }\textbf {\bibinfo {volume} {19}},\ \bibinfo {pages} {842} (\bibinfo {year} {2014})}\BibitemShut {NoStop}%
\bibitem [{\citenamefont {Oliveira}\ \emph {et~al.}(2018)\citenamefont {Oliveira}, \citenamefont {Chan},\ and\ \citenamefont {Leonel}}]{oliveira2018scaling}%
  \BibitemOpen
  \bibfield  {author} {\bibinfo {author} {\bibfnamefont {D.~F.}\ \bibnamefont {Oliveira}}, \bibinfo {author} {\bibfnamefont {K.~S.}\ \bibnamefont {Chan}},\ and\ \bibinfo {author} {\bibfnamefont {E.~D.}\ \bibnamefont {Leonel}},\ }\bibfield  {title} {\bibinfo {title} {Scaling invariance in a social network with limited attention and innovation},\ }\href@noop {} {\bibfield  {journal} {\bibinfo  {journal} {Physics Letters A}\ }\textbf {\bibinfo {volume} {382}},\ \bibinfo {pages} {3376} (\bibinfo {year} {2018})}\BibitemShut {NoStop}%
\bibitem [{\citenamefont {Oliveira}\ and\ \citenamefont {Robnik}(2012)}]{oliveira2012scaling}%
  \BibitemOpen
  \bibfield  {author} {\bibinfo {author} {\bibfnamefont {D.~F.}\ \bibnamefont {Oliveira}}\ and\ \bibinfo {author} {\bibfnamefont {M.}~\bibnamefont {Robnik}},\ }\bibfield  {title} {\bibinfo {title} {Scaling invariance in a time-dependent elliptical billiard},\ }\href@noop {} {\bibfield  {journal} {\bibinfo  {journal} {International Journal of Bifurcation and Chaos}\ }\textbf {\bibinfo {volume} {22}},\ \bibinfo {pages} {1250207} (\bibinfo {year} {2012})}\BibitemShut {NoStop}%
\bibitem [{\citenamefont {Livorati}\ \emph {et~al.}(2011)\citenamefont {Livorati}, \citenamefont {Loskutov},\ and\ \citenamefont {Leonel}}]{livorati2011family}%
  \BibitemOpen
  \bibfield  {author} {\bibinfo {author} {\bibfnamefont {A.~L.}\ \bibnamefont {Livorati}}, \bibinfo {author} {\bibfnamefont {A.}~\bibnamefont {Loskutov}},\ and\ \bibinfo {author} {\bibfnamefont {E.~D.}\ \bibnamefont {Leonel}},\ }\bibfield  {title} {\bibinfo {title} {A family of stadium-like billiards with parabolic boundaries under scaling analysis},\ }\href@noop {} {\bibfield  {journal} {\bibinfo  {journal} {Journal of Physics A: Mathematical and Theoretical}\ }\textbf {\bibinfo {volume} {44}},\ \bibinfo {pages} {175102} (\bibinfo {year} {2011})}\BibitemShut {NoStop}%
\bibitem [{\citenamefont {Chernov}\ and\ \citenamefont {Markarian}(2006)}]{chernov2006chaotic}%
  \BibitemOpen
  \bibfield  {author} {\bibinfo {author} {\bibfnamefont {N.}~\bibnamefont {Chernov}}\ and\ \bibinfo {author} {\bibfnamefont {R.}~\bibnamefont {Markarian}},\ }\href@noop {} {\emph {\bibinfo {title} {Chaotic billiards}}},\ \bibinfo {number} {127}\ (\bibinfo  {publisher} {American Mathematical Soc.},\ \bibinfo {year} {2006})\BibitemShut {NoStop}%
\bibitem [{\citenamefont {Berry}(1981)}]{berry1981regularity}%
  \BibitemOpen
  \bibfield  {author} {\bibinfo {author} {\bibfnamefont {M.~V.}\ \bibnamefont {Berry}},\ }\bibfield  {title} {\bibinfo {title} {Regularity and chaos in classical mechanics, illustrated by three deformations of a circular'billiard'},\ }\href@noop {} {\bibfield  {journal} {\bibinfo  {journal} {European Journal of Physics}\ }\textbf {\bibinfo {volume} {2}},\ \bibinfo {pages} {91} (\bibinfo {year} {1981})}\BibitemShut {NoStop}%
\bibitem [{\citenamefont {Bunimovich}\ and\ \citenamefont {Dettmann}(2005)}]{bunimovich2005open}%
  \BibitemOpen
  \bibfield  {author} {\bibinfo {author} {\bibfnamefont {L.}~\bibnamefont {Bunimovich}}\ and\ \bibinfo {author} {\bibfnamefont {C.}~\bibnamefont {Dettmann}},\ }\bibfield  {title} {\bibinfo {title} {Open circular billiards and the riemann hypothesis},\ }\href@noop {} {\bibfield  {journal} {\bibinfo  {journal} {Physical review letters}\ }\textbf {\bibinfo {volume} {94}},\ \bibinfo {pages} {100201} (\bibinfo {year} {2005})}\BibitemShut {NoStop}%
\bibitem [{\citenamefont {Koiller}\ \emph {et~al.}(1996)\citenamefont {Koiller}, \citenamefont {Markarian}, \citenamefont {Oliffson~Kamphorst},\ and\ \citenamefont {Pinto~de Carvalho}}]{koiller1996static}%
  \BibitemOpen
  \bibfield  {author} {\bibinfo {author} {\bibfnamefont {J.}~\bibnamefont {Koiller}}, \bibinfo {author} {\bibfnamefont {R.}~\bibnamefont {Markarian}}, \bibinfo {author} {\bibfnamefont {S.}~\bibnamefont {Oliffson~Kamphorst}},\ and\ \bibinfo {author} {\bibfnamefont {S.}~\bibnamefont {Pinto~de Carvalho}},\ }\bibfield  {title} {\bibinfo {title} {Static and time-dependent perturbations of the classical elliptical billiard},\ }\href@noop {} {\bibfield  {journal} {\bibinfo  {journal} {Journal of Statistical Physics}\ }\textbf {\bibinfo {volume} {83}},\ \bibinfo {pages} {127} (\bibinfo {year} {1996})}\BibitemShut {NoStop}%
\bibitem [{\citenamefont {Lenz}\ \emph {et~al.}(2008)\citenamefont {Lenz}, \citenamefont {Diakonos},\ and\ \citenamefont {Schmelcher}}]{lenz2008tunable}%
  \BibitemOpen
  \bibfield  {author} {\bibinfo {author} {\bibfnamefont {F.}~\bibnamefont {Lenz}}, \bibinfo {author} {\bibfnamefont {F.~K.}\ \bibnamefont {Diakonos}},\ and\ \bibinfo {author} {\bibfnamefont {P.}~\bibnamefont {Schmelcher}},\ }\bibfield  {title} {\bibinfo {title} {Tunable fermi acceleration in the driven elliptical billiard},\ }\href@noop {} {\bibfield  {journal} {\bibinfo  {journal} {Physical review letters}\ }\textbf {\bibinfo {volume} {100}},\ \bibinfo {pages} {014103} (\bibinfo {year} {2008})}\BibitemShut {NoStop}%
\bibitem [{\citenamefont {Lopac}\ \emph {et~al.}(2002)\citenamefont {Lopac}, \citenamefont {Mrkonji{\'c}},\ and\ \citenamefont {Radi{\'c}}}]{lopac2002chaotic}%
  \BibitemOpen
  \bibfield  {author} {\bibinfo {author} {\bibfnamefont {V.}~\bibnamefont {Lopac}}, \bibinfo {author} {\bibfnamefont {I.}~\bibnamefont {Mrkonji{\'c}}},\ and\ \bibinfo {author} {\bibfnamefont {D.}~\bibnamefont {Radi{\'c}}},\ }\bibfield  {title} {\bibinfo {title} {Chaotic dynamics and orbit stability in the parabolic oval billiard},\ }\href@noop {} {\bibfield  {journal} {\bibinfo  {journal} {Physical Review E}\ }\textbf {\bibinfo {volume} {66}},\ \bibinfo {pages} {036202} (\bibinfo {year} {2002})}\BibitemShut {NoStop}%
\bibitem [{\citenamefont {Leonel}\ \emph {et~al.}(2009)\citenamefont {Leonel}, \citenamefont {Oliveira},\ and\ \citenamefont {Loskutov}}]{leonel2009fermi}%
  \BibitemOpen
  \bibfield  {author} {\bibinfo {author} {\bibfnamefont {E.~D.}\ \bibnamefont {Leonel}}, \bibinfo {author} {\bibfnamefont {D.~F.}\ \bibnamefont {Oliveira}},\ and\ \bibinfo {author} {\bibfnamefont {A.}~\bibnamefont {Loskutov}},\ }\bibfield  {title} {\bibinfo {title} {Fermi acceleration and scaling properties of a time dependent oval billiard},\ }\href@noop {} {\bibfield  {journal} {\bibinfo  {journal} {Chaos: An Interdisciplinary Journal of Nonlinear Science}\ }\textbf {\bibinfo {volume} {19}} (\bibinfo {year} {2009})}\BibitemShut {NoStop}%
\bibitem [{\citenamefont {Tomsovic}\ and\ \citenamefont {Heller}(1993)}]{tomsovic1993long}%
  \BibitemOpen
  \bibfield  {author} {\bibinfo {author} {\bibfnamefont {S.}~\bibnamefont {Tomsovic}}\ and\ \bibinfo {author} {\bibfnamefont {E.~J.}\ \bibnamefont {Heller}},\ }\bibfield  {title} {\bibinfo {title} {Long-time semiclassical dynamics of chaos: The stadium billiard},\ }\href@noop {} {\bibfield  {journal} {\bibinfo  {journal} {Physical Review E}\ }\textbf {\bibinfo {volume} {47}},\ \bibinfo {pages} {282} (\bibinfo {year} {1993})}\BibitemShut {NoStop}%
\bibitem [{\citenamefont {Dettmann}\ and\ \citenamefont {Georgiou}(2009)}]{dettmann2009survival}%
  \BibitemOpen
  \bibfield  {author} {\bibinfo {author} {\bibfnamefont {C.~P.}\ \bibnamefont {Dettmann}}\ and\ \bibinfo {author} {\bibfnamefont {O.}~\bibnamefont {Georgiou}},\ }\bibfield  {title} {\bibinfo {title} {Survival probability for the stadium billiard},\ }\href@noop {} {\bibfield  {journal} {\bibinfo  {journal} {Physica D: Nonlinear Phenomena}\ }\textbf {\bibinfo {volume} {238}},\ \bibinfo {pages} {2395} (\bibinfo {year} {2009})}\BibitemShut {NoStop}%
\bibitem [{\citenamefont {Lozej}\ and\ \citenamefont {Robnik}(2018)}]{lozej2018aspects}%
  \BibitemOpen
  \bibfield  {author} {\bibinfo {author} {\bibfnamefont {{\v{C}}.}~\bibnamefont {Lozej}}\ and\ \bibinfo {author} {\bibfnamefont {M.}~\bibnamefont {Robnik}},\ }\bibfield  {title} {\bibinfo {title} {Aspects of diffusion in the stadium billiard},\ }\href@noop {} {\bibfield  {journal} {\bibinfo  {journal} {Physical Review E}\ }\textbf {\bibinfo {volume} {97}},\ \bibinfo {pages} {012206} (\bibinfo {year} {2018})}\BibitemShut {NoStop}%
\bibitem [{\citenamefont {Lichtenberg}\ and\ \citenamefont {Lieberman}(2013)}]{lichtenberg2013regular}%
  \BibitemOpen
  \bibfield  {author} {\bibinfo {author} {\bibfnamefont {A.~J.}\ \bibnamefont {Lichtenberg}}\ and\ \bibinfo {author} {\bibfnamefont {M.~A.}\ \bibnamefont {Lieberman}},\ }\href@noop {} {\emph {\bibinfo {title} {Regular and chaotic dynamics}}},\ Vol.~\bibinfo {volume} {38}\ (\bibinfo  {publisher} {Springer Science \& Business Media},\ \bibinfo {year} {2013})\BibitemShut {NoStop}%
\bibitem [{\citenamefont {Bohrer}\ and\ \citenamefont {Xiao}(2020)}]{bohrer2020complex}%
  \BibitemOpen
  \bibfield  {author} {\bibinfo {author} {\bibfnamefont {C.~H.}\ \bibnamefont {Bohrer}}\ and\ \bibinfo {author} {\bibfnamefont {J.}~\bibnamefont {Xiao}},\ }\bibfield  {title} {\bibinfo {title} {Complex diffusion in bacteria},\ }\href@noop {} {\bibfield  {journal} {\bibinfo  {journal} {Physical Microbiology}\ ,\ \bibinfo {pages} {15}} (\bibinfo {year} {2020})}\BibitemShut {NoStop}%
\bibitem [{\citenamefont {Xu}(2021)}]{xu2021global}%
  \BibitemOpen
  \bibfield  {author} {\bibinfo {author} {\bibfnamefont {Z.}~\bibnamefont {Xu}},\ }\bibfield  {title} {\bibinfo {title} {On the global attractivity of a nonlocal and vector-bias malaria model},\ }\href@noop {} {\bibfield  {journal} {\bibinfo  {journal} {Applied Mathematics Letters}\ }\textbf {\bibinfo {volume} {121}},\ \bibinfo {pages} {107459} (\bibinfo {year} {2021})}\BibitemShut {NoStop}%
\bibitem [{\citenamefont {Lu}\ \emph {et~al.}(2019)\citenamefont {Lu}, \citenamefont {Lu},\ and\ \citenamefont {He}}]{lu2019modeling}%
  \BibitemOpen
  \bibfield  {author} {\bibinfo {author} {\bibfnamefont {H.}~\bibnamefont {Lu}}, \bibinfo {author} {\bibfnamefont {J.}~\bibnamefont {Lu}},\ and\ \bibinfo {author} {\bibfnamefont {L.}~\bibnamefont {He}},\ }\bibfield  {title} {\bibinfo {title} {Modeling and estimation of pollen-mediated gene flow at the landscape scale},\ }\href@noop {} {\bibfield  {journal} {\bibinfo  {journal} {Ecological Indicators}\ }\textbf {\bibinfo {volume} {106}},\ \bibinfo {pages} {105500} (\bibinfo {year} {2019})}\BibitemShut {NoStop}%
\bibitem [{\citenamefont {Gibert}\ and\ \citenamefont {Yeakel}(2019)}]{gibert2019laplacian}%
  \BibitemOpen
  \bibfield  {author} {\bibinfo {author} {\bibfnamefont {J.~P.}\ \bibnamefont {Gibert}}\ and\ \bibinfo {author} {\bibfnamefont {J.~D.}\ \bibnamefont {Yeakel}},\ }\bibfield  {title} {\bibinfo {title} {Laplacian matrices and turing bifurcations: Revisiting levin 1974 and the consequences of spatial structure and movement for ecological dynamics},\ }\href@noop {} {\bibfield  {journal} {\bibinfo  {journal} {Theoretical Ecology}\ }\textbf {\bibinfo {volume} {12}},\ \bibinfo {pages} {265} (\bibinfo {year} {2019})}\BibitemShut {NoStop}%
\bibitem [{\citenamefont {Chkhetiani}\ and\ \citenamefont {Golitsyn}(2014)}]{chkhetiani2014detection}%
  \BibitemOpen
  \bibfield  {author} {\bibinfo {author} {\bibfnamefont {O.}~\bibnamefont {Chkhetiani}}\ and\ \bibinfo {author} {\bibfnamefont {G.}~\bibnamefont {Golitsyn}},\ }\bibfield  {title} {\bibinfo {title} {Detection and dispersion of diffusion tracer spots and their life times},\ }in\ \href@noop {} {\emph {\bibinfo {booktitle} {Doklady Mathematics}}},\ Vol.~\bibinfo {volume} {89}\ (\bibinfo {organization} {Pleiades Publishing},\ \bibinfo {year} {2014})\ pp.\ \bibinfo {pages} {245--249}\BibitemShut {NoStop}%
\bibitem [{\citenamefont {Loskutov}\ \emph {et~al.}(2000)\citenamefont {Loskutov}, \citenamefont {Ryabov},\ and\ \citenamefont {Akinshin}}]{loskutov2000properties}%
  \BibitemOpen
  \bibfield  {author} {\bibinfo {author} {\bibfnamefont {A.}~\bibnamefont {Loskutov}}, \bibinfo {author} {\bibfnamefont {A.}~\bibnamefont {Ryabov}},\ and\ \bibinfo {author} {\bibfnamefont {L.}~\bibnamefont {Akinshin}},\ }\bibfield  {title} {\bibinfo {title} {Properties of some chaotic billiards with time-dependent boundaries},\ }\href@noop {} {\bibfield  {journal} {\bibinfo  {journal} {Journal of Physics A: Mathematical and General}\ }\textbf {\bibinfo {volume} {33}},\ \bibinfo {pages} {7973} (\bibinfo {year} {2000})}\BibitemShut {NoStop}%
\bibitem [{\citenamefont {Loskutov}\ \emph {et~al.}(1999)\citenamefont {Loskutov}, \citenamefont {Ryabov},\ and\ \citenamefont {Akinshin}}]{loskutov1999mechanism}%
  \BibitemOpen
  \bibfield  {author} {\bibinfo {author} {\bibfnamefont {A.~Y.}\ \bibnamefont {Loskutov}}, \bibinfo {author} {\bibfnamefont {A.}~\bibnamefont {Ryabov}},\ and\ \bibinfo {author} {\bibfnamefont {L.}~\bibnamefont {Akinshin}},\ }\bibfield  {title} {\bibinfo {title} {Mechanism of fermi acceleration in dispersing billiards with time-dependent boundaries},\ }\href@noop {} {\bibfield  {journal} {\bibinfo  {journal} {Journal of Experimental and Theoretical Physics}\ }\textbf {\bibinfo {volume} {89}},\ \bibinfo {pages} {966} (\bibinfo {year} {1999})}\BibitemShut {NoStop}%
\bibitem [{\citenamefont {Oliveira}\ and\ \citenamefont {Leonel}(2010)}]{oliveira2010suppressing}%
  \BibitemOpen
  \bibfield  {author} {\bibinfo {author} {\bibfnamefont {D.~F.}\ \bibnamefont {Oliveira}}\ and\ \bibinfo {author} {\bibfnamefont {E.~D.}\ \bibnamefont {Leonel}},\ }\bibfield  {title} {\bibinfo {title} {Suppressing fermi acceleration in a two-dimensional non-integrable time-dependent oval-shaped billiard with inelastic collisions},\ }\href@noop {} {\bibfield  {journal} {\bibinfo  {journal} {Physica A: Statistical Mechanics and its Applications}\ }\textbf {\bibinfo {volume} {389}},\ \bibinfo {pages} {1009} (\bibinfo {year} {2010})}\BibitemShut {NoStop}%
\bibitem [{\citenamefont {Oliveira}\ \emph {et~al.}(2011)\citenamefont {Oliveira}, \citenamefont {Vollmer},\ and\ \citenamefont {Leonel}}]{oliveira2011fermi}%
  \BibitemOpen
  \bibfield  {author} {\bibinfo {author} {\bibfnamefont {D.~F.}\ \bibnamefont {Oliveira}}, \bibinfo {author} {\bibfnamefont {J.}~\bibnamefont {Vollmer}},\ and\ \bibinfo {author} {\bibfnamefont {E.~D.}\ \bibnamefont {Leonel}},\ }\bibfield  {title} {\bibinfo {title} {Fermi acceleration and its suppression in a time-dependent lorentz gas},\ }\href@noop {} {\bibfield  {journal} {\bibinfo  {journal} {Physica D: Nonlinear Phenomena}\ }\textbf {\bibinfo {volume} {240}},\ \bibinfo {pages} {389} (\bibinfo {year} {2011})}\BibitemShut {NoStop}%
\bibitem [{\citenamefont {Batisti{\'c}}\ and\ \citenamefont {Robnik}(2011)}]{batistic2011fermi}%
  \BibitemOpen
  \bibfield  {author} {\bibinfo {author} {\bibfnamefont {B.}~\bibnamefont {Batisti{\'c}}}\ and\ \bibinfo {author} {\bibfnamefont {M.}~\bibnamefont {Robnik}},\ }\bibfield  {title} {\bibinfo {title} {Fermi acceleration in time-dependent billiards: theory of the velocity diffusion in conformally breathing fully chaotic billiards},\ }\href@noop {} {\bibfield  {journal} {\bibinfo  {journal} {Journal of Physics A: Mathematical and Theoretical}\ }\textbf {\bibinfo {volume} {44}},\ \bibinfo {pages} {365101} (\bibinfo {year} {2011})}\BibitemShut {NoStop}%
\bibitem [{\citenamefont {Batisti{\'c}}(2014)}]{batistic2014exponential}%
  \BibitemOpen
  \bibfield  {author} {\bibinfo {author} {\bibfnamefont {B.}~\bibnamefont {Batisti{\'c}}},\ }\bibfield  {title} {\bibinfo {title} {Exponential fermi acceleration in general time-dependent billiards},\ }\href@noop {} {\bibfield  {journal} {\bibinfo  {journal} {Physical Review E}\ }\textbf {\bibinfo {volume} {90}},\ \bibinfo {pages} {032909} (\bibinfo {year} {2014})}\BibitemShut {NoStop}%
\bibitem [{\citenamefont {Balakrishnan}(2008)}]{balakrishnan2008elements}%
  \BibitemOpen
  \bibfield  {author} {\bibinfo {author} {\bibfnamefont {V.}~\bibnamefont {Balakrishnan}},\ }\href@noop {} {\emph {\bibinfo {title} {Elements of nonequilibrium statistical mechanics}}},\ Vol.~\bibinfo {volume} {3}\ (\bibinfo  {publisher} {Springer},\ \bibinfo {year} {2008})\BibitemShut {NoStop}%
\end{thebibliography}%
%

\end{document}